\documentstyle[emulate_apj,apjfonts,epsf]{article}
\gdef\h50min{$h_{50}^{-1}$}
\gdef\1054{MS\,1054$-$03}
\gdef\kms{km\,s$^{-1}$}
\lefthead{van Dokkum et al.}
\righthead{CM relation in MS\,1054--03}
\slugcomment{Accepted for publication in the Astrophysical Journal}
\begin{document}
\title{HST Photometry and Keck Spectroscopy of
the Rich Cluster MS\,1054--03:
Morphologies, Butcher-Oemler Effect and the Color-Magnitude Relation
at $z=0.83$ \altaffilmark{1,2}}
\author{
Pieter G. van Dokkum\altaffilmark{3} and Marijn Franx}
\affil{Leiden Observatory, P.O. Box 9513, NL-2300 RA, Leiden, The Netherlands}
\author{Daniel Fabricant}
\affil{Harvard-Smithsonian Center for Astrophysics, 60 Garden Street,
Cambridge, MA 02318}
\author{Garth D. Illingworth}
\affil{University of California Observatories/Lick Observatory,
University of California, Santa Cruz, CA 95064}
\and
\author{Daniel D. Kelson}
\affil{D. T. M., Carnegie Institution
of Washington, 5241 Broad Branch Road, NW, Washington DC, 20015}

\altaffiltext{1}
{Based on observations with the NASA/ESA {\em Hubble Space
Telescope}, obtained at the Space Telescope Science Institute, which
is operated by AURA, Inc., under NASA contract NAS 5--26555.}
\altaffiltext{2}
{Based on observations obtained at the W.\ M.\ Keck Observatory,
which is operated jointly by the California Institute of
Technology and the University of California.}
\altaffiltext{3}
{Present address:
California Institute of Technology, MS\,105-24, Pasadena, CA 91125}

\begin{abstract}

We present a study of 81 $I$-band selected, spectroscopically-confirmed
members of the X-ray cluster \1054{} at $z=0.83$. Redshifts and
spectral types were determined from Keck spectroscopy.  Morphologies
and accurate colors were determined from a large mosaic of HST WFPC2
images in $R_{F606W}$ and $I_{F814W}$, corresponding to
$U$ and $B$ in the restframe.  Early-type galaxies constitute only $44$\,\% of
this galaxy population. This fraction is much lower than in comparable
rich clusters at low redshift. Thirty-nine percent are spiral
galaxies, and 17\,\% are mergers.  The early-type galaxies follow a
tight and well-defined color-magnitude relation, with the exception of
a few outliers.  The observed scatter is $0.029 \pm 0.005$ magnitudes
in restframe $U-B$.  Most of the mergers lie close to the CM relation
defined by the early-type galaxies. They are bluer by only $0.07 \pm
0.02$ magnitudes, and the scatter in their colors is $0.07 \pm
0.04$ magnitudes. Spiral galaxies in \1054{} exhibit a large range in their
colors.  The bluest spiral galaxies are $\sim 0.7$ magnitudes bluer
than the early-type galaxies, but the majority is within $\pm 0.2$
magnitudes of the early-type galaxy sequence. The red
colors of the mergers and the majority of the spiral galaxies are
reflected in the fairly low Butcher-Oemler blue fraction of \1054{}:
$f_B=0.22 \pm 0.05$, similar to intermediate redshift clusters and
much lower than previously reported values for clusters at $z \sim
0.8$. The slope and scatter of the CM relation of early-type galaxies
are roughly constant with redshift, confirming previous studies that
were based on ground-based color measurements and very limited
membership information.  However, the scatter in the combined sample
of early-type galaxies and mergers (i.e., the sample of {\em future}
early-type galaxies) is twice as high as the scatter of the early-type
galaxies alone. This is a direct demonstration of the ``progenitor
bias'': high redshift early-type galaxies seem to form a homogeneous,
old population because the progenitors of the youngest present-day
early-type galaxies are not included in the sample.

\end{abstract}

\section{Introduction}

The star formation history of early-type galaxies provides an
important constraint on models for galaxy formation and a test for
stellar population synthesis models. Because present-day
early-type galaxies consist mostly of old stars, the evolution of their
stellar populations may provide clues to conditions in the early
Universe.
Early-type galaxies (elliptical galaxies and S0 galaxies)
are very efficiently studied in clusters.  They
form the dominant population in nearby rich clusters (Oemler 1974;
Dressler 1980),
and most of our knowledge of the formation and evolution of early-type
galaxies has come from studies of clusters at $0<z<0.5$.

Strong constraints on the star formation histories of early-type
galaxies have been derived from the redshift
evolution of the Fundamental
Plane relation (van Dokkum \& Franx 1996, Kelson et al.\
1997, Pahre 1998, Bender et al.\ 1998,
van Dokkum et al.\ 1998b, J\o{}rgensen et al.\ 1999,
Kelson et al.\
2000) and the color-magnitude relation  (e.g., Ellis et al.\ 1997,
Stanford, Eisenhardt, \& Dickinson 1998, van Dokkum et al.\ 1998a,
Kodama et al.\ 1998).
The strongest constraints on the {\em mean} star
formation epoch are provided by the evolution of
the Fundamental Plane, mainly because 
the mean luminosity evolution is large compared to the mean
color evolution.
van Dokkum et al.\ (1998b) find a slow luminosity evolution of
early-type galaxies to $z=0.83$, and conclude
that their stars were formed at
$z>2.8$ for a Salpeter
(1955) IMF, $\Omega_m=0.3$ and $\Omega_{\Lambda}=0$, or $z>1.7$ for
$\Omega_m=0.3$ and $\Omega_{\Lambda}=0.7$.

Whereas the Fundamental Plane can be used to measure the mean
luminosity evolution of early-type galaxies with high accuracy, it is
difficult to obtain a sufficiently large sample to study the evolution of
its slope and scatter (but not impossible, see Kelson et al.\
2000). Furthermore, the Fundamental Plane can only be applied to
early-type galaxies, and concentrating on the early-type galaxy
population alone may lead to an incomplete picture of their evolution.
There is growing evidence that a significant
fraction of present-day early-type galaxies was assembled relatively
recently (Dressler et al.\
1997; van Dokkum et al.\ 1999).  They may have formed from gas
stripping of spiral galaxies (e.g.,
Gunn \& Gott 1972, Tytler \& Vidal 1978, Solanes \& Salvador-Sole 1992,
Couch et al.\ 1998, Abadi, Moore, \& Bower 1999),
through tidal interactions (e.g., Fried 1988, Lavery, Pierce, \&
McClure 1992, Oemler, Dressler, \& Butcher 1997, Moore, Lake, \&
Katz 1998, Couch et al.\ 1998), or in
mergers (e.g., Lavery et al.\ 1992, van Dokkum et al.\ 1999).
Early-type galaxies in high redshift clusters may therefore
form a special subset of the progenitors of early-type galaxies in
nearby clusters.  Failure to consider the other progenitors may result
in biased age estimates (see, e.g., Franx \& van Dokkum 1996).

The evolution of the color-magnitude (CM) relation can provide
important constraints additional to those provided by the Fundamental
Plane. The CM relation is useful for
studying all morphological types and large photometric samples
can be obtained straightforwardly.
The combination of the mean color evolution of
galaxies and their luminosity evolution can constrain the IMF
(e.g., Kelson et al.\ 2000). The evolution of the slope
of the CM relation is sensitive to
age differences between high mass galaxies and low mass galaxies (see,
e.g., Kodama et al.\ 1998). The scatter in the CM relation is
particularly important. It can be measured to high accuracy
with HST and it constrains the age spread of
galaxies of a given luminosity (Bower, Lucey, \& Ellis 1992;
Ellis et al.\ 1997; van Dokkum et al.\ 1998a; Stanford et al.\ 1998).
The scatter is proportional to $\Delta \ln (\tau)
= \Delta \tau / \langle \tau \rangle$, where $\tau$ is the
luminosity-weighted age (see van Dokkum et al.\ 1998a).

The evolution of the color-magnitude relation has been studied
extensively (e.g., Aragon-Salamanca et al.\ 1993,
Stanford, Eisenhardt, \& Dickinson 1995,
Ellis et al.\ 1997, Stanford et al.\ 1998,
Kodama et al.\ 1998). These studies
are in remarkable agreement: the slope and scatter seem to
be roughly constant with redshift, and the mean color evolution
is consistent with passive evolution of an old stellar population
that was formed at high redshift.
However, these studies suffer from limited membership information and
areal coverage (usually one HST+WFPC2 pointing). Because the field
contamination among late-type galaxies becomes very large at $z \sim 1$
usually only early-type galaxies are considered, and even these
morphologically selected samples suffer from significant
contamination. For example, Stanford et al.\ (1998) estimate that the
field contamination among the early-type galaxies
in their $z>0.6$ clusters is $20 -
40$\,\%, depending on the richness of the cluster. An additional
complication is that the true contamination is expected to vary
considerably on the small angular scales of the fields
used for these studies.

We are undertaking a program to obtain large field HST+WFPC2 mosaics of
clusters at $0.3 < z < 1$, combined with extensive spectroscopy from the
ground. This program is complimentary to other HST studies of distant
clusters, in the sense that we observe only a few clusters, but over a
large field and with redshifts for $\gtrsim 200$ (field and cluster)
galaxies for each cluster field. The CM relation in the
cluster CL\,1358+62 at $z=0.33$ was presented by van Dokkum
et al.\ (1998a). We used a sample of 194 confirmed members
observed with HST. It was found that S0
galaxies in the outskirts of the cluster
show a much larger scatter in their colors and are bluer on average than
those in the central regions, providing evidence for recent infall of
galaxies from the field. Here, we present extensive Keck spectroscopy
and a large HST WFPC2 mosaic in two filters of the cluster \1054{} at
$z=0.83$. Redshifts, morphologies and accurate colors are presented for
a sample of 81 confirmed cluster members. One of
the most striking results of
our survey is the large number of merging systems in \1054{} (van Dokkum
et al.\ 1999). In the present study we focus on the CM relation of
confirmed members. The aims are to test whether the early-types are a
homogeneous population at $z \approx 1$, and to compare their CM
relation to those of other morphological types.

\section{Data}
\subsection{Spectroscopy}

The spectroscopic survey of the \1054{} field used the LRIS spectrograph
(Oke et al.\ 1995) on the 10\,m W.\,M.\ Keck II Telescope.
The primary purpose of the survey was to obtain a sample
of $I$-band selected, spectroscopically confirmed cluster galaxies.

\subsubsection{Sample Selection and Observations}
 
Targets for the spectroscopy were selected from a photometric catalog
of objects in the \1054{} field. The catalog was created from a 900\,s
Keck $I$-band image of the cluster, centered on the Brightest Cluster
Galaxy (BCG), and spanning $5' \times 7'$. The seeing in the image is
$1\farcs0$ FWHM. Objects were detected with FOCAS (Valdes et al.\
1995). $I$ band magnitudes were measured through $3\arcsec$ diameter
apertures. The 412 objects in the catalog with $20.0 < I < 22.7$ were
selected as potential targets for the spectroscopy.  The $I$ magnitude
of the BCG equals $20.2$.  Six slitmasks were designed, with $\sim 40$
slitlets each. Objects in the magnitude range $20.0 < I < 22.2$ were
given the highest priority. Objects located outside the boundaries of
our HST mosaic were given low priority irrespective of their
magnitude.  The positions and position angles of the masks maximize
the total number of slitlets within the area covered by our HST
imaging.

The \1054{} field was observed 1998 February 28 and March 1.  All six
slitmasks were exposed, using the 400 mm$^{-1}$ grating blazed
at 8500\,\AA, and $1\farcs2$ wide slits. Five masks were exposed for
$2\times 1200\,$s, and one mask for 1200\,s. The seeing was $\approx
0\farcs 8$ during the observations.
The instrumental resolution, as measured from sky lines, is $\approx
7$\,\AA{} FWHM at $7500$\,\AA{}.  The wavelength coverage depends on
the position of each slitlet in the multislit mask, but is typically
$5700$ -- $9500$\,\AA. The [O\,{\sc ii}] 3727\,\AA{} doublet and the
$4000$\,\AA{} break redshifted to $z=0.83$ are covered in all
slitlets. The data reduction is described in
Appendix \ref{specred.sec}.

\subsubsection{Redshifts and Completeness}
\label{complete.sec}

A total of 256 1D spectra were extracted from the 2D data. This
number includes serendipitous detections.
We calculated the sampling rate
by dividing the number of objects in the spectroscopic sample by
the number of objects in the photometric catalog.
Figure \ref{complete.plot}(a) shows the
sampling rate as a function of magnitude, calculated in $\pm 0.25$
magnitude bins around each galaxy. Approximately 90\,\% of
all $I \approx 21$ objects within the boundaries of our
HST mosaic were observed. The sampling rate drops
precipitously at $I \gtrsim 22$, consistent with our selection
process.


\vbox{
\begin{center}
\leavevmode
\hbox{%
\epsfxsize=7cm
\epsffile{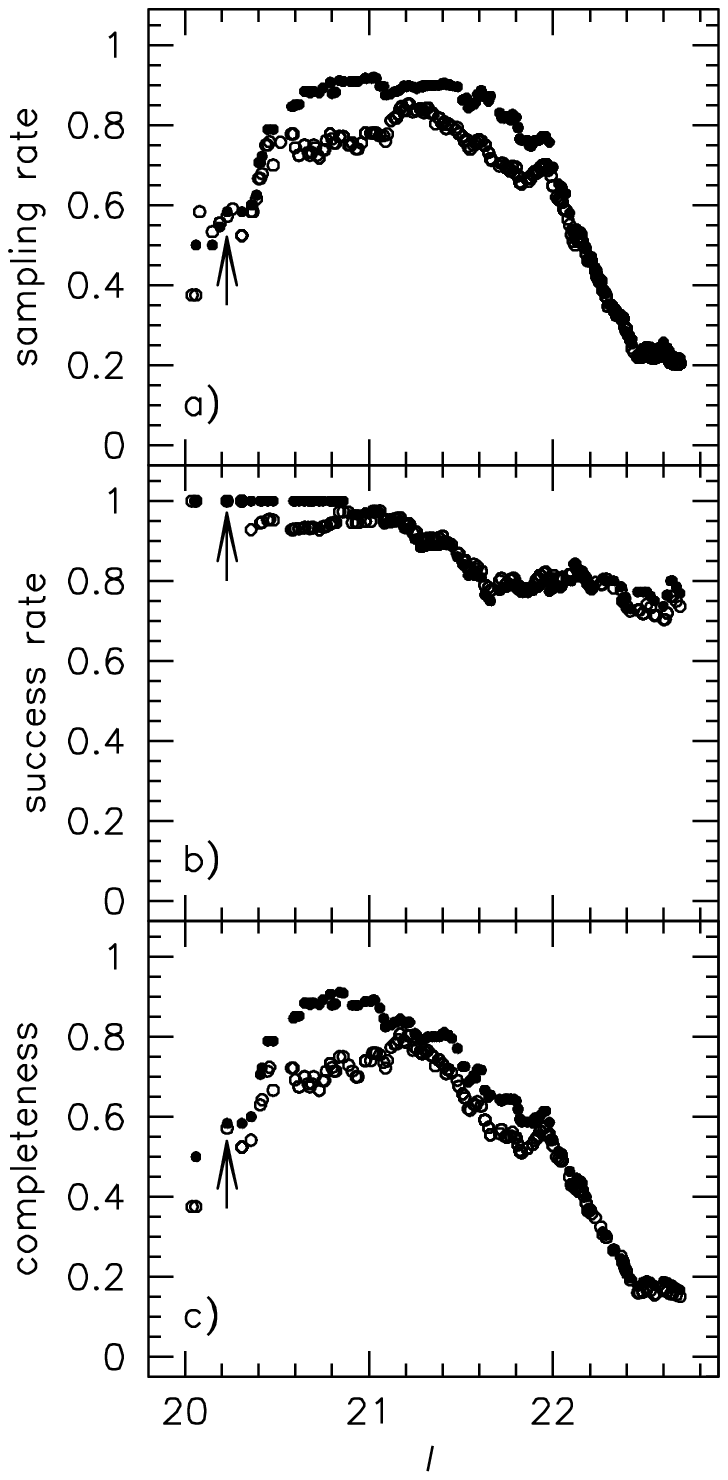}}
\begin{small}
\figcaption{\label{complete.plot} \small
(a) The sampling rate as a function of magnitude,
calculated by dividing the number of galaxies in our spectroscopic
sample by
the number of galaxies in the photometric catalog, in $\pm 0.25$ mag
bins centered on each galaxy. Open symbols are for the full sample,
solid symbols for objects covered by our HST imaging. The arrow
indicates the BCG. (b) The
success rate of measuring the redshift of observed objects.
The success rate is still $\approx 80$\,\% at $I
= 22.7$. (c) The completeness,
calculated by dividing the number of galaxies with measured redshift
by the number of galaxies in the photometric catalog. At a given
magnitude, the completeness
is the product of the sampling rate and the success rate.
The incompleteness at the faint end of the sample
is primarily caused by sparse sampling, and not by the inability to
measure redshifts of observed galaxies.
}
\end{small}
\end{center}}

Redshifts were measured from emission lines using
{\sc splot} in IRAF, or determined from absorption lines using the
cross-correlation routine {\sc xcsao}. For
the cross-correlations two template spectra were used:
the spectrum of the nearby early-type galaxy NGC 7331, and the spectrum
of the E+A galaxy 209 in CL\,1358+62 (Fisher et al.\ 1998). The
galaxies and the templates were cross-correlated
in the 3750 -- 4500\,\AA{} restframe wavelength range.
Redshifts were determined for 200 objects. Ten turned out to be
stars, and four were observed twice. Therefore, the number of
galaxies with redshifts is 186. The success rate of measuring the
redshift of observed objects is shown in Fig.\ \ref{complete.plot}(b).
The success rate is still as high as $\approx 80$\,\% at $I=22.7$.

We investigate the completeness of the spectroscopic sample by comparing
the number of galaxies with measured redshifts to the number of galaxies
in the photometric catalog. Figure\ \ref{complete.plot}(c) shows the
completeness (or the ``magnitude selection function'', cf.\
Yee et al.\ 1996) as a function of $I$ magnitude.
The sample is $65$\,\% complete for
$20.0 < I < 22.2$, and $16$\,\% complete for $22.2 \leq I < 22.7$.
Within the boundaries of the HST mosaic the completeness is $73$\,\%
for $20.0 < I < 22.2$, and $17$\,\% for $22.2 \leq I < 22.7$.
As is often the case in this type of survey (e.g., Lilly 1993, Fisher
et al.\ 1998) the incompleteness at the faint end of the sample is
primarily caused by sparse sampling, and not by the inability to
measure redshifts of observed galaxies.

An important question is whether there is a color bias
in the sample of galaxies with redshifts.
In Fig.\ \ref{color.plot}(a)
the distribution of $R-I$ colors is shown for the full photometric
sample, in the magnitude range $22.2 \leq I < 22.7$.
The $R$
magnitudes were determined from an image kindly provided by G.\ Luppino
(Luppino \& Kaiser 1997); the $R-I$ colors were normalized
to the colors of the BCG.
Figure \ref{color.plot}(b) shows the color distribution
for the sample of galaxies with redshifts (solid line). As a result
of the sparse sampling at faint magnitudes there is
a small bias against blue galaxies in the spectroscopic sample.


\vbox{
\begin{center}
\leavevmode
\hbox{%
\epsfxsize=8cm
\epsffile{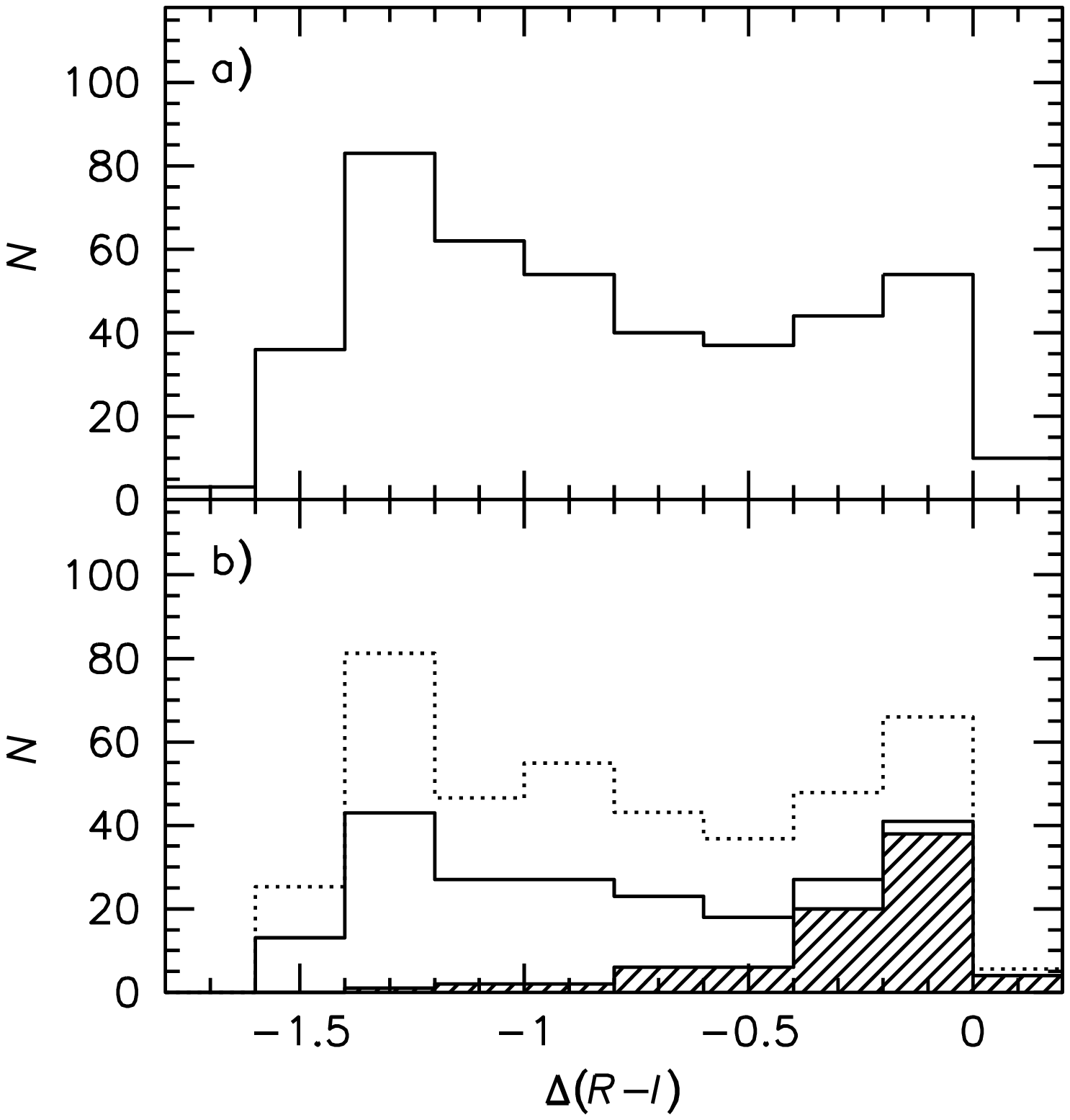}}
\begin{small}
\figcaption{\small
(a) Color distribution of all galaxies in the photometric catalog.
(b) Color distribution of galaxies with a measured redshift (solid
histogram).
The broken histogram is the same distribution, corrected for
incompleteness (see text).
The hatched histogram shows the color distribution
of confirmed cluster members.
\label{color.plot}}
\end{small}
\end{center}}

We test whether there is an additional source of
color bias in the spectroscopic sample by assigning each galaxy
a weight equal to the inverse of the magnitude selection function
shown in Fig.\ \ref{complete.plot}(c).
The resulting distribution is shown by the broken line in
Fig.\ \ref{color.plot}(b), and is very similar to the
color distribution of the full photometric sample shown in
Fig.\ \ref{color.plot}(a). We conclude that there is no bias against blue
galaxies in the spectroscopic sample after correcting for the sparse
sampling at faint magnitudes. We note that the effect of this
completeness correction is small for the cluster galaxies.
The hatched histogram in Fig.\ \ref{color.plot}(b) shows the
color distribution of confirmed cluster galaxies. Among blue
galaxies with measured redshifts the fraction
of cluster members is very low.
We will return to this issue in Sect.\ \ref{bo.sec}.

\subsubsection{Redshift Distribution and Velocity Dispersion}

Figure \ref{zhis.plot} shows the
redshift distribution of our sample of 186 galaxies.
The peak at $z=0.83$ is conspicuous, and
\1054{} is well separated from the field
in redshift space.  In the redshift range $0.78 < z < 0.87$ all galaxies
have redshifts in the interval $0.8132<z<0.847$, i.e., within $\pm
2.7\,\sigma_{\rm cl}$ (Tran et al.\ 1999).
The separation of cluster galaxies and
field galaxies is therefore straightforward.


\vbox{
\begin{center}
\leavevmode
\hbox{%
\epsfxsize=8cm
\epsffile{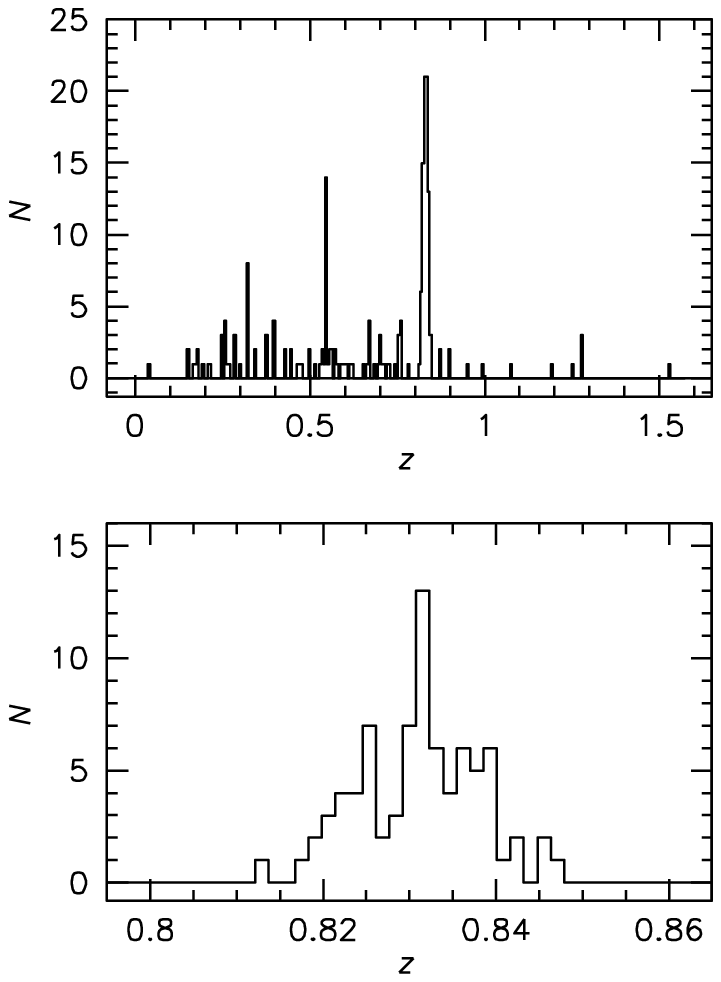}}
\begin{small}
\figcaption{
\small
Redshift distribution of galaxies in our spectroscopic sample.
The peak at $z=0.83$ is conspicuous, and well separated from the field.
We find at least two peaks at lower redshifts.
\label{zhis.plot}}
\end{small}
\end{center}}

There are 78 cluster members in our spectroscopic sample.  Two
additional cluster members were observed serendipitously, in slits
pointed at other objects. These galaxies with $I \sim 24$ will not be
considered in subsequent analysis. Redshifts of cluster members are
listed in Table 1 in Appendix A. Uncertainties in radial velocity are
typically $\sim 30$\,\kms{} in the restframe ($\Delta z \sim 0.0002$).
We can refine the Donahue et al.\ (1998) and Tran et al.\ (1999)
determinations of the mean redshift and velocity dispersion of
\1054{}.  Using the biweight estimators for location and scale (Beers,
Flynn, \& Gebhardt 1990) we find $\langle z \rangle = 0.8315 \pm
0.0007$ and $\sigma = 1150 \pm 97$\,\kms{} for our sample of 78
members.  The uncertainty in $\sigma$ was determined from bootstrap
resampling.

\subsubsection{Spectral Types}

We determined spectral types of the cluster galaxies from the
strengths of the [O\,{\sc ii}]\,3727\,\AA{} emission line and the
H$\delta$ 4102\,\AA{} Balmer line. The procedure is described in
Appendix \ref{lines.sec}.  In our restframe $B$ selected sample of 78
cluster galaxies, there are 52 absorption line galaxies (67\,\%), 15
emission line galaxies (19\,\%), and 11 E+A galaxies (14\,\%). These
fractions are somewhat uncertain due to the large number of galaxies
which are less than $1\sigma$ removed from other spectral types.  The
number fractions of the various spectral types are not significantly
different from those in intermediate redshift clusters (e.g., Couch \&
Sharples 1987, Balogh et al.\ 1998, Fisher et al.\ 1998, Poggianti et
al.\ 1999). The fraction of emission line galaxies in \1054{} is
significantly lower than that of the optically selected massive
cluster CL\,1604+4304 at $z=0.89$, as determined by Postman, Lubin, \&
Oke (1998). These authors find that 11 out of 22 cluster galaxies have
EW [O\,{\sc ii}]\,$>15$\,\AA. In contrast, only 7 out of 78 galaxies
in \1054{} have EW [O\,{\sc ii}]\,$>15$\,\AA. We note, however, that
the Postman et al.\ (1998) sample extends to fainter magnitudes than
our sample. Furthermore, our sample was selected in restframe $B$ and
the Postman et al.\ (1998) sample in restframe $U$.  Finally,
CL\,1604+4304 is an optically selected cluster whereas \1054{} was
selected because of its high X-ray luminosity.

The emission line galaxies in
\1054{} are less concentrated toward the center of the cluster than the
other galaxies.  The fraction of emission line galaxies is $6$\,\%
within a radius of $0.5$\,\h50min\,Mpc, and $31$\,\% outside
$R=0.5$\,\h50min\,Mpc. These numbers are similar to those found for low
and intermediate redshift clusters (e.g., de Theije \& Katgert 1999,
Balogh et al.\ 1998).

We combine the new redshifts with
samples of Donahue et al.\ (1998) and Tran et al.\ (1999),
giving a total number of 89 spectroscopically confirmed cluster
members. For subsequent analysis, we use the 81 confirmed
members that are covered by our HST imaging.

\subsection{HST WFPC2 Imaging}
\label{imaging.sec}

\begin{figure*}[p]
\epsfxsize=17cm
\epsffile{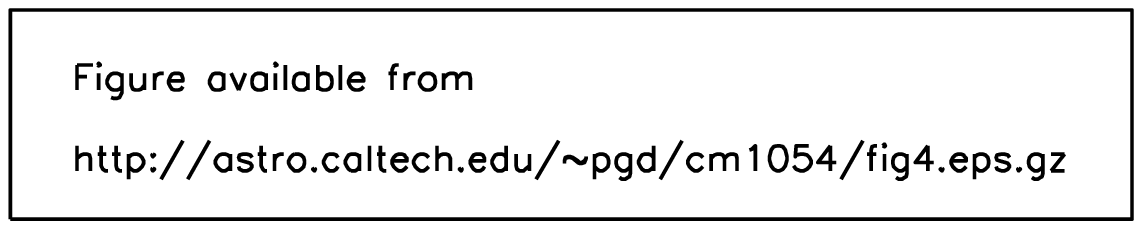}
\caption{\small
Hubble Space Telescope WFPC2 mosaic of the cluster \1054{}
at $z=0.83$. This color image was created from the F606W and F814W
exposures, and spans roughly $4' \times 6'$ ($2 \times 3$\,\h50min\,Mpc).
The red cluster galaxies are conspicuous. Note their irregular
and elongated distribution.
\label{colfield.plot}}
\end{figure*}

We have obtained a large HST WFPC2 mosaic of \1054, consisting of six
slightly overlapping pointings.
The area of the mosaic is $25$\,arcmin$^2$, or
8\,$h_{50}^{-2}$\,Mpc$^2$.
\1054{} was observed with the F606W and F814W filters on 1998 May 30.
Exposure times were 6500\,s in each
passband and at each position.  The images were interlaced to improved
the sampling by a factor $\sqrt{2}$. Interlacing leaves the pixel
values intact, and there is no loss of information.  The reduction and
interlacing procedures are described in Appendix B.  A color image of
the mosaic is shown in Fig.\ \ref{colfield.plot}. The red cluster
galaxies stand out, and the irregular structure of the cluster is
obvious.

Restframe $B_z$ magnitudes and accurate $(U-B)_z$ colors 
for the eighty-one cluster members were determined from the
HST mosaic, as described in Appendix B.
Magnitudes and colors of these galaxies are listed in Table
2 in Appendix B.


\section{Morphologies}

\subsection{Visual Classifications}

The $F814W$ WFPC2 images of the 81 confirmed members in the HST mosaic
were visually classified
by three of us (PGvD, MF, and DF). Details of the classification
procedure can be found in Fabricant, Franx, \& van Dokkum (2000).
In summary,
the images were compared to images of nearby galaxies from the survey by
Jansen et al.\ (2000). The classification system is identical to that of
Fabricant et al.\ (2000), and similar to the system of Smail et
al.\ (1997). In order to extract as much information as possible from
our data the galaxies were displayed at both the original
$0\farcs1$ WFC resolution, and the
$0\farcs071$ resolution of the interlaced images.
The three sets of classifications were compared to assess
their internal errors. In general, classifications agreed within the broad
categories of early-type galaxy, spiral galaxy,
or merger. If galaxy types are
binned in the two categories early-types (E, E/S0, S0) and
other types the three classifiers agreed on 76\,\% of
all galaxies. To combine the classifications
we used the same combination rules as Fabricant et al.\ (2000).
The resulting classifications are listed in Table 2 in Appendix B.

\subsection{The Early-Type Galaxy Fraction}
\label{morph.sec}

The relative fractions of elliptical galaxies, S0 galaxies,
spiral galaxies, and mergers
are 22\,\%, 22\,\%, 39\,\%, and 17\,\% respectively, where
E/S0 galaxies and S0/a galaxies
are evenly split between their neighboring types.
The ratio of
elliptical galaxies to S0 galaxies is $\sim 1$.  This quantity is somewhat
uncertain as it is very difficult to distinguish
elliptical galaxies from S0 galaxies (see Sect.\ \ref{evoearly.sec}).
The fraction of early-type galaxies (E+E/S0+S0)
is usually better determined (e.g., Fabricant et al.\ 2000). Only 34
out of 77 classified galaxies (44\,\%) are early-type galaxies.
This number includes 50\,\% of galaxies classified as S0/a.

This low early-type fraction is surprising, because the projected
galaxy density of \1054{} is very high.  We calculated the projected
galaxy density following the prescription of Dressler (1980). The
average galaxy density $\log \rho_{\rm proj} \approx
1.5$\,$h_{50}^{-2}$\,Mpc$^{-2}$ within a circular aperture of radius
$120\arcsec$ centered on the Brightest Cluster Galaxy (BCG).  This
aperture contains $87$\,\% of our 81 confirmed members.  The
logarithm of the average projected density rises to $\approx
1.8$\,$h_{50}^{-2}$\,Mpc$^{-2}$ in the central $60\arcsec$
($570$\,\h50min\,kpc).
At low redshift, early-type galaxies constitute $\sim 80$\,\%
of the galaxy population in
regions with such high galaxy densities (Dressler 1980),
much higher than the value of 44\,\% we find
for \1054{} at $z=0.83$.

In Fig.\ \ref{earlyrad.plot} we test whether the fraction of early-type
galaxies in \1054{} is a function of $R$, the projected distance from
the BCG.  The histogram shows the average early-type fraction in
0.5\,\h50min\,Mpc wide bins. The early-type fraction is constant at $R
\lesssim 1$\,\h50min\,Mpc, and may decline outside this radius.  There
is no rise of the early-type fraction in the central regions of the
cluster, contrary to what is seen in low redshift clusters (Dressler
1980).


\1054{} has an irregular and elongated appearance (see Hoekstra, Franx,
\& Kuijken 2000 and Fig.\ \ref{colfield.plot}).
We test whether the low early-type fraction in the
central regions of \1054{} is related to substructure, by calculating
the early-type fraction within apertures centered on the three most
luminous galaxies in the cluster. Their positions coincide with the
three strongest peaks in the luminosity density distribution.  The
early-type fractions in the immediate vicinity of these three galaxies
are in fact somewhat lower than the average fraction in the central
region of the cluster: they are 33\,\%, 38\,\% and 38\,\% within
circles with radii $R = 0.25$\,\h50min\,Mpc centered on galaxies 1484,
1325 and 1405 respectively.  We conclude that even in the highest
density regions of the cluster the fraction of early-type galaxies
does not exceed $50$\,\%.

\vbox{
\begin{center}
\leavevmode
\hbox{%
\epsfxsize=7cm
\epsffile{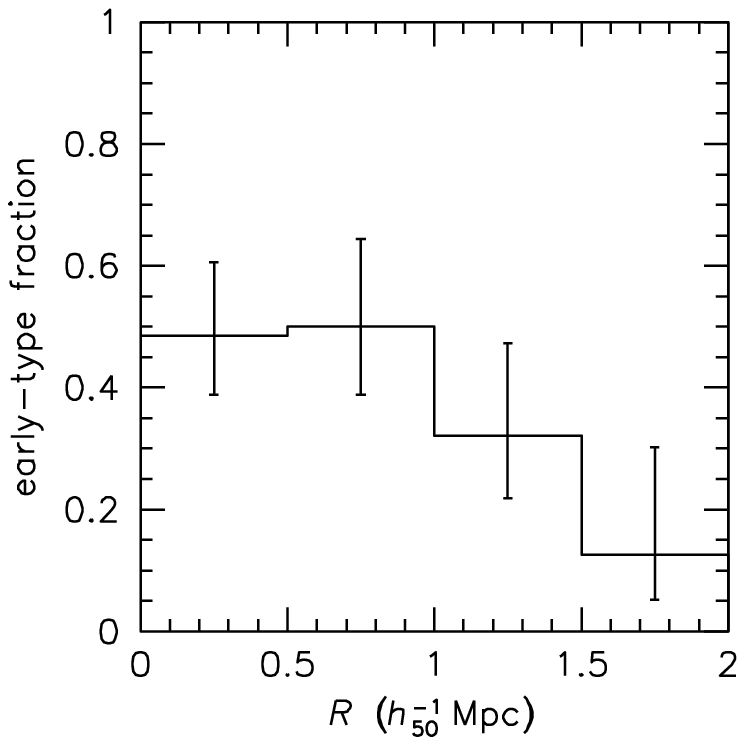}}
\begin{small}
\figcaption{\small
Early-type fraction plotted against $R$, the projected
distance from the BCG. The histogram shows the average early-type
fraction in
0.5\,\h50min\,Mpc wide bins. The early-type fraction is low, even in the
central regions of the cluster.
\label{earlyrad.plot}}
\end{small}
\end{center}}

\subsection{The Early-Type Fraction in Rich Clusters from $z=0$
to $z=1$}
\label{evoearly.sec}

We can compare the fraction of early-type galaxies in \1054{} to
the fractions in other clusters. The evolution of the early-type
fraction is shown in Fig.\ \ref{evoearly.plot}.
The large
symbols are CL1358+62 at $z=0.33$ and \1054{} at $z=0.83$.
For CL1358+62 we used the sample of confirmed members presented
in Fisher et al.\ (1998) and van Dokkum et al.\ (1998a),
with morphological classifications from Fabricant et al.\ (2000).
For consistency
with Dressler et al.\ (1997) we computed the early-type fraction
in these clusters within the physical radius that corresponds to
one WFPC2 pointing at $z=0.45$ ($R=600$\,\h50min\,kpc). The star
symbol at $z=0.04$ represents the concentrated clusters
from Dressler (1980), taking the revised morphological fractions
from Dressler et al.\ (1997). Triangles
are from Andreon, Davoust, \& Heim (1997) and small circles
from Dressler et al.\ (1997). The connected symbols
at $z=0.41$ are for CL0939+47, which was classified
both by Andreon et al.\ and Dressler et al. 
The cross is CL1358+62 at $z=0.33$,
as classified by A.\ Dressler (see Fabricant et al.\
2000). The square at $z=0.90$ is derived from twelve confirmed
members of the cluster CL1604+43 (Lubin et al.\ 1998).
Filled symbols indicate clusters
with $L_X>10^{44.5}\,h_{50}^{-2}$\,ergs\,s$^{-1}$. X-ray
luminosities are taken from Fabricant, McClintock, \& Bautz
(1991), Smail et al.\ (1997), Donahue et al.\ (1998),
and Postman et al.\ (1998).


There is a clear trend with redshift in Fig.\
\ref{evoearly.plot}: high redshift clusters have lower early-type
fractions than low redshift clusters.
This result is not very sensitive to the details of the
classifications. In those cases where clusters were classified
by two different groups the fractions are very similar.
The early-type fraction is much more stable from classifier
to classifier than the ratio between the number
of elliptical galaxies and the number of S0 galaxies
(for discussions of the robustness of E/S0 fractions see
Smail et al.\ 1997, Andreon 1998, Postman 1999,
and Fabricant et al.\ 2000). This may not be surprising;
as demonstrated by work on nearby
galaxies the distinction between ellipticals and S0s is very difficult
to make (e.g., Rix \& White 1990, J\o{}rgensen \& Franx 1994).

\vbox{
\begin{center}
\leavevmode
\hbox{%
\epsfxsize=8cm
\epsffile{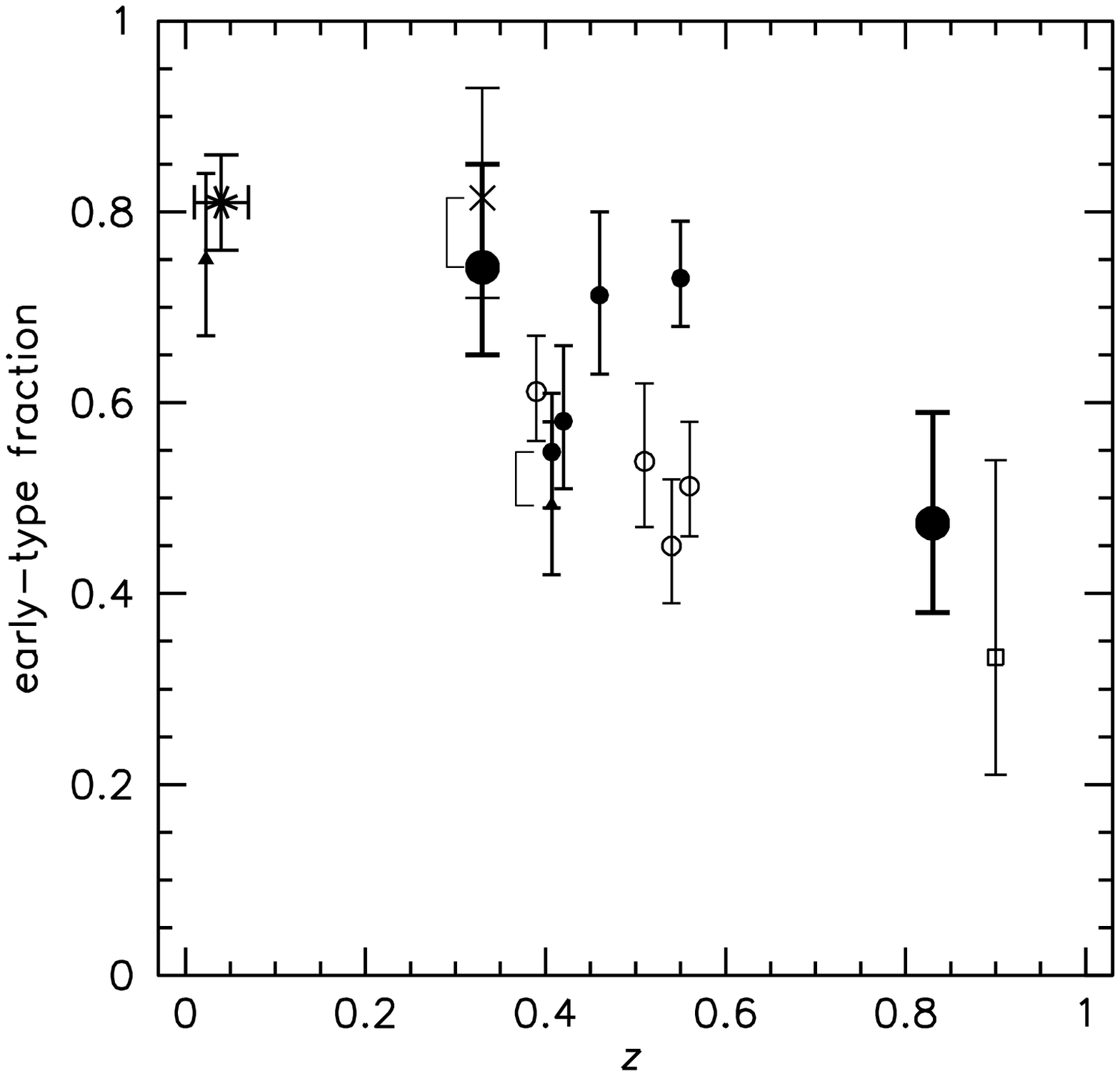}}
\begin{small}
\figcaption{\small
Evolution of the early-type fraction in rich clusters. The large
symbols are CL1358+62 at $z=0.33$ and \1054{} at $z=0.83$.
Sources of literature data are Andreon et al.\ (1997) (triangles),
Dressler (1980) (star),
Dressler et al.\ (1997) (small circles), and Lubin et al.\ (1998)
(square). Filled symbols indicate clusters
with $L_X>10^{44.5}\,h_{50}^{-2}$\,ergs\,s$^{-1}$.
Two clusters (CL1358+62 at $z=0.33$ and CL0939+47 at $z=0.41$)
have been classified twice, by independent groups; note the good agreement
between these independent classifications (see text for details).
The early-type fraction is a function of redshift: the early-type
fraction at $z=0$ is roughly twice as large as the early-type fraction
at $z\sim 1$.
\label{evoearly.plot}}
\end{small}
\end{center}}

\begin{figure*}[b]
\epsfxsize=15cm
\epsffile{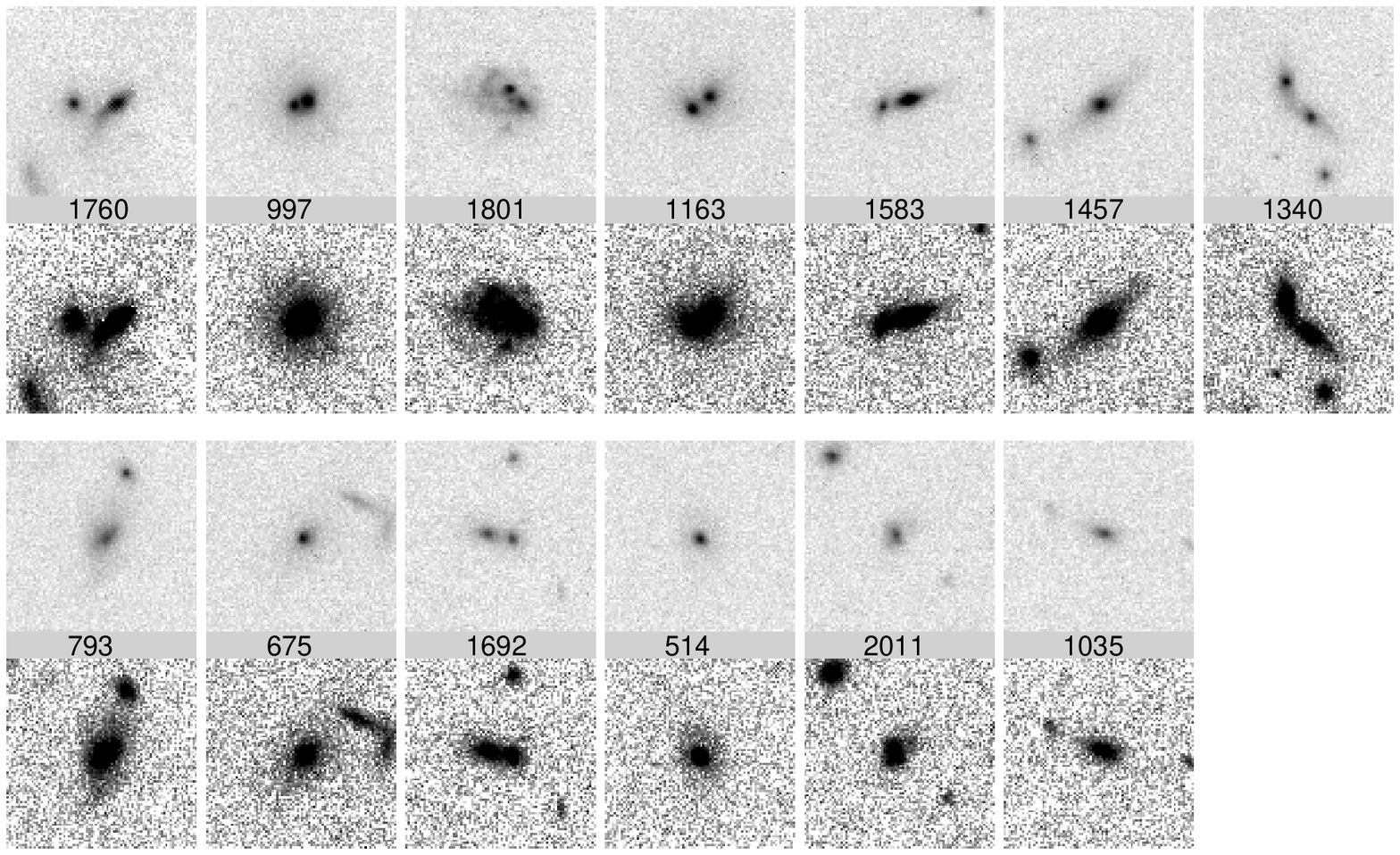}
\caption{\small
F814W images of the thirteen galaxies classified as ``merger/peculiar'',
displayed at two contrast levels. The mergers show a variety of
features; we find double nuclei, tidal tails and/or interacting pairs with
distorted morphologies. Note that all mergers are
confirmed members of the cluster.
\label{mergers.plot}}
\end{figure*}

The strong evolution of the early-type fraction implies that $\sim
50$\,\% of early-type galaxies in present-day rich clusters were
either accreted onto the cluster or transformed from other galaxy
types between $z=0$ and $z=1$. Transformations (or a combination of
transformations and accretion) are the most probable mechanism,
because it is unlikely that the fraction of early-type galaxies among
accreted galaxies exceeds the fraction of early-type galaxies among
galaxies that are already in the cluster.

\section{Mergers}

One of the most surprising results of our survey is the high fraction
of galaxies classified as ``merger/peculiar'' (van Dokkum et al.\
1999).  Thirteen out of 77 classified galaxies are mergers.
F814W images of all mergers are shown in Fig.\
\ref{mergers.plot}. The resolution of these images is
$0\farcs{}07$\,pixel$^{-1}$ (see Sect.\ \ref{interlace.sec}).
The mergers display a variety of features. We
find interacting pairs with distorted morphologies (e.g., 1340),
double nuclei (e.g., 997) and galaxies with strong tidal features
(e.g., 1760).  Note that all mergers, including the interacting pairs
and double nuclei, are counted as single objects.


\begin{figure*}[t]
\epsfxsize=17cm
\epsffile{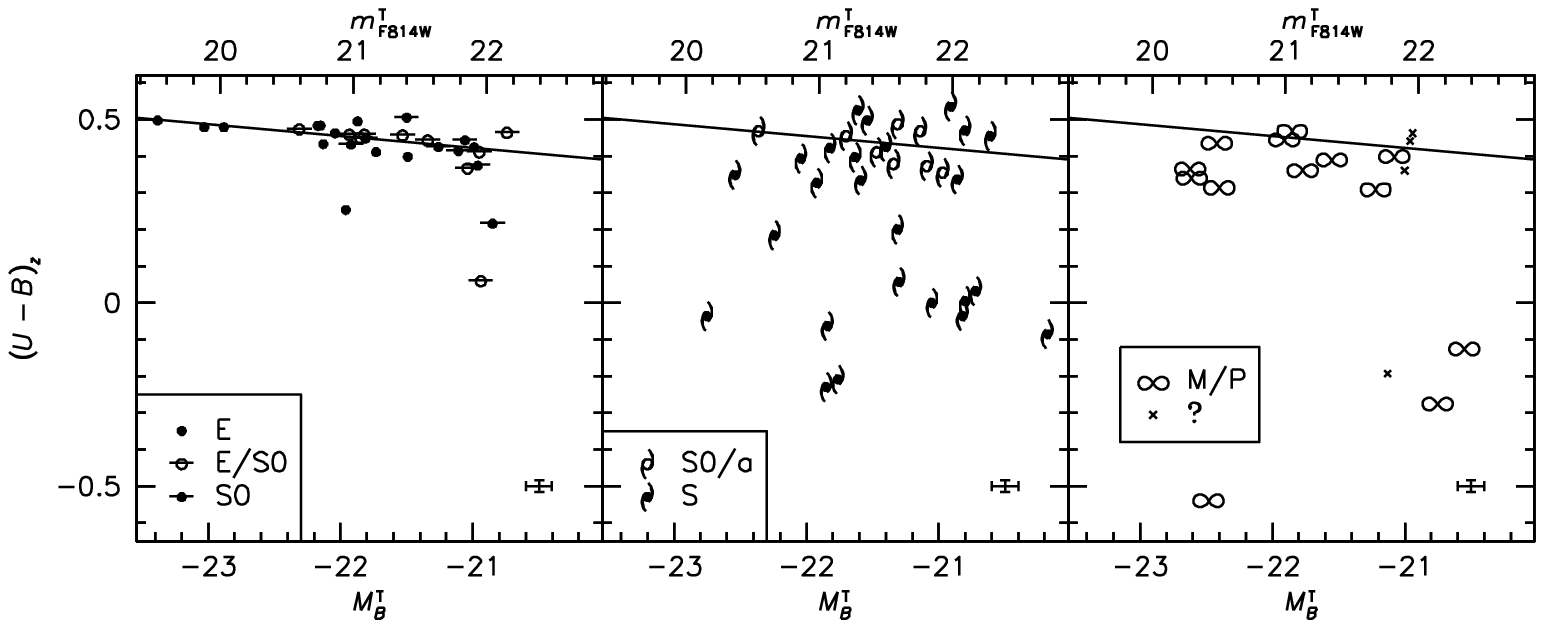}
\caption{\small
The restframe $U-B$ versus $B$
color-magnitude relation for different morphologies.
There is a clear trend with morphology. Most early-type galaxies
follow a tight and well-defined CM relation, although there
are a few outliers.
The drawn line in each plot is a fit to the CM relation of the
early-type galaxies (Es, E/S0s, and S0s).
Spiral galaxies have a large scatter, and are on average bluer than
early-type galaxies in \1054. The mergers follow the same CM relation as
the early-type galaxies, but are slightly offset to the blue
and show a larger scatter.
\label{cm_4.plot}}
\end{figure*}

The mergers are very luminous: the 13 mergers have a median luminosity
$M_B^T \approx -22$ ($\sim 2 L_*$ at $z=0.83$), and five of the
sixteen most luminous cluster galaxies were classified as mergers. We
calculated the luminosity of merging pairs by taking the sum of the
luminosities of the individual merging galaxies. As argued in
van Dokkum et al.\ (1999), the majority of the
mergers will probably evolve into elliptical galaxies. The merger
fraction in MS\,1054 is comparable to the elliptical fraction: the
number of elliptical galaxies ``in formation'' is similar to the
number of elliptical galaxies already formed. Assuming the galaxy
population in MS\,1054--03 is typical for rich clusters at its
redshift, this implies up to $\sim 50$\,\% of elliptical galaxies
formed in mergers at $z<1$ (see van Dokkum et al.\ 1999).

\section{The Color-Magnitude Relation and the Butcher-Oemler Effect}

\subsection{The CM Relation for Different Morphologies}

The color-magnitude relation for spectroscopically confirmed cluster
members is shown in Fig.\ \ref{cm_4.plot}. There is a strong trend
with morphology. Most early-type galaxies follow a tight and well
defined relation, but there are three early-type galaxies which are fairly
blue. Luminous blue early-type
galaxies are very rare in
clusters at lower redshift (e.g., van Dokkum et al.\ 1998a). The
spiral galaxies have a much larger scatter in their colors than the
early-type galaxies. As might be expected, some are very blue,
but a fairly large number are
quite red and lie close to the relation of the early-type
galaxies. The
mergers are only slightly bluer than the early-type galaxies, and follow a
much tighter relation than the spiral galaxies.


We quantified these effects by
determining the form of the CM relation
from a fit to the early-type galaxies,
and subtracting this fit from the observed
relation. For each morphological subsample we then
computed the offset and scatter in the CM relation from the distribution
of residuals. The fitting procedure minimizes the scatter in the
residuals, computed using the biweight statistic (Beers et al.\
1990). The biweight gives low weight to outliers,
and was also used in the studies of van
Dokkum et al.\ (1998a) and Stanford et al.\
(1998). The fit to the early-type galaxies
(Es, E/S0s, and S0s) has the form
\begin{equation}
(U-B)_z = -0.032 (M_B^T+22) + 0.454.
\end{equation}
Note that no early-type galaxies were excluded from the fit.
We tested the sensitivity of our results to the fitting procedure
by comparing the biweight minimalisation to a least squares fit excluding the
three bluest early-type galaxies. The least squares fit
has the form $(U-B)_z = -0.030 (M_B^T + 22) + 0.456$, and gives very
similar results for the scatter of the CM relation as the biweight fit.
The predicted colors from the CM relation were subtracted from the
observed colors. For each morphological subsample
the mean and scatter in the distribution of residuals
$\Delta(U-B)_z$ were computed using the biweight statistic.
Uncertainties in the scatter were
determined from bootstrap resampling. The results are listed in Table 3.

The observed scatter in the early-type galaxies is low at $0.029 \pm
0.005$. The intrinsic scatter is $0.024 \pm 0.006$. This scatter is much
smaller than that of the full sample ($0.081 \pm 0.026$).
The sample of early-type galaxies is too small to divide it into
smaller subsamples.

From Table 3 we can see that
the spiral galaxies and mergers are more heterogeneous in their colors
than the early-type galaxies, with the spirals showing the largest
scatter.
On average the spiral galaxies are much bluer than the
early-type galaxies, but not all of them are blue: in fact,
only 12 out of 26 spiral galaxies and none of the S0/a galaxies
are more than 0.2 magnitudes bluer than the CM relation.
The mergers seem to follow the same CM relation as the early-type
galaxies,
but are slightly offset to the blue and show a larger scatter.
The average offset between the CM relation of the early-type
galaxies
and the mergers is $-0.07 \pm 0.02$ magnitudes.
The scatter in the CM relation of the mergers is $0.07 \pm 0.04$,
more than twice as large as the scatter of the early-type
galaxies, and
comparable to the mean offset from the CM relation.

\subsection{Radial Dependence of the Color-Magnitude Relation}

\begin{figure*}[t]
\epsfxsize=17cm
\epsffile{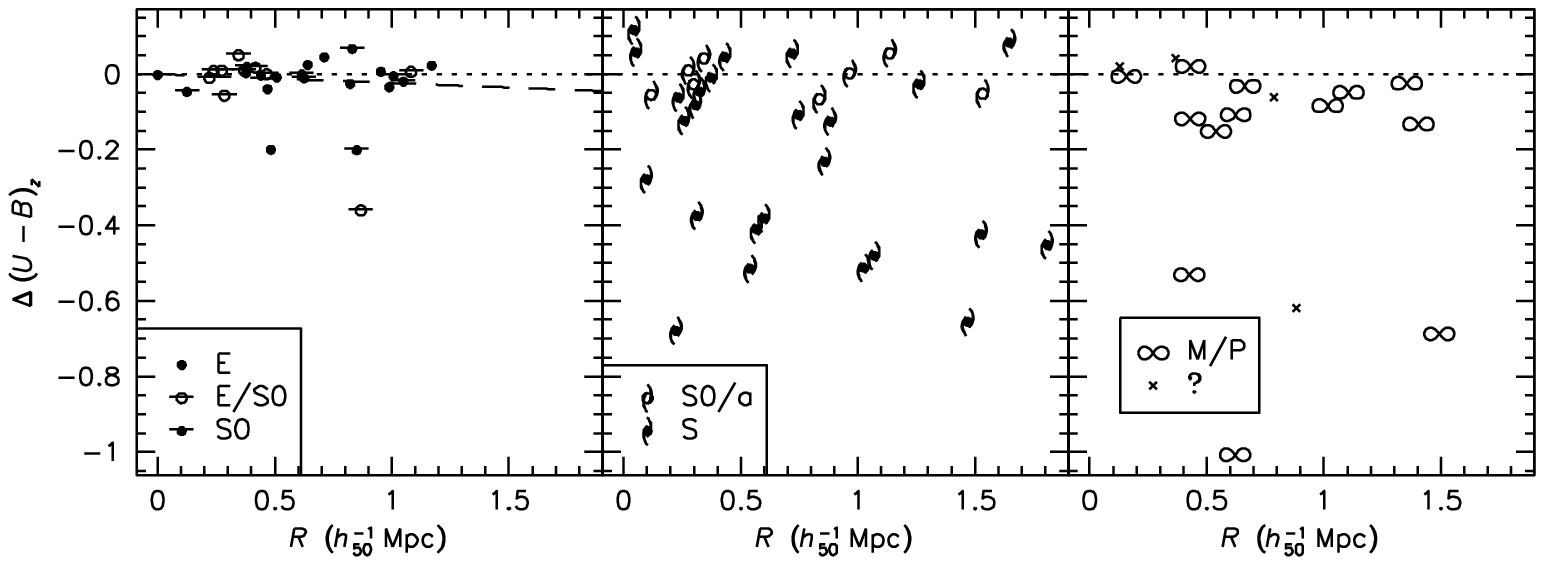}
\caption{\small
Residual from the color-magnitude relation plotted against $R$, the
projected distance from the BCG. The dashed line indicates the trend
found in the CM relation of S0 galaxies
in the cluster CL\,1358+62 at $z=0.33$.
There is no significant correlation between
$(U-B)_z$ and $R$ for the early-type galaxies, although we cannot
exclude a similar trend as in CL\,1358+62. Note the absence of
early-type galaxies at $R>1.2$\,\h50min\,Mpc.
\label{radres.plot}}
\end{figure*}

In the cluster CL\,1358+62 at $z=0.33$, the scatter and offset of the
color-magnitude relation are dependent on $R$, the projected distance to
the BCG; S0 galaxies in the outer parts of CL\,1358+62 have
a larger scatter in the CM relation and a bluer mean color than S0
galaxies in
the inner parts (van Dokkum et al.\ 1998a). If this trend is caused by an age gradient in the
cluster one might expect it to strengthen at higher redshift.
In Fig.\ \ref{radres.plot} the residuals
from the CM relation are plotted against $R$.
There is no significant correlation between $\Delta (U-B)_z$ and $R$
for the early-type galaxies. The trend observed in CL\,1358+62 is
indicated by the broken line in Fig.\ \ref{radres.plot}; our data are
not of sufficient quality to detect or rule
out such a trend in \1054{}.


We note that there seems to be a larger
fraction of red spiral galaxies in the core of the cluster than at larger
radii, although the numbers are small.
Only three out of eleven spiral galaxies at $R<0.5$\,\h50min\,Mpc
are more than 0.2 magnitudes bluer than the CM relation,
compared to nine out of fifteen at $R>0.5$\,\h50min\,Mpc.

\subsection{The Butcher-Oemler Effect}
\label{bo.sec}

There have been very few attempts to
extend measurements of the Butcher \& Oemler (1978a, 1984)
effect beyond $z \sim 0.5$,
partly because field contamination is more severe at higher
redshift. Here, we use our extensive dataset on \1054{} to
measure the Butcher-Oemler effect in a cluster at $z=0.83$.

Butcher \& Oemler (1984) defined the blue fraction as the fraction of
the cluster population within $R=R_{30}$ that is more than 0.2
magnitudes bluer in restframe $B-V$ than the ridge line of the CM
relation, with $R_{30}$ the radius of a circular aperture containing
30\,\% of the cluster galaxies (Butcher \& Oemler 1984). They also
imposed a fixed magnitude cutoff of $M_V<-20$ in the restframe.
Note that this procedure ignores luminosity evolution.

The value of $R_{30}$ can be determined from the projected surface
density profile of the cluster (cf.\ Butcher \& Oemler 1978b).
We determine the surface
density profile of \1054{} from our sample of spectroscopically confirmed
members, after correcting for incompleteness
(see Sect.\ \ref{complete.sec}).
Butcher \& Oemler (1984) used unpublished models by Aarseth to
extrapolate the observed surface density profile to large radii.
At large radii, these models are well approximated by a power
law (see Butcher \& Oemler 1978b). From a power law fit to the
observed surface density profile of \1054{} we find $\approx 20$\,\%
of its galaxy population is outside the field of our spectroscopic
survey. We find $R_{30}=49 \arcsec$, and a concentration index
$C \equiv \log (R_{60}/R_{20}) = 0.53$. We conclude that \1054{} is a
compact cluster according to the Butcher \& Oemler (1984) criterion.

To test the robustness of our approach we also
determine the surface density profile from red galaxies
in the HST mosaic, by selecting all galaxies
to $I_{F814W}=24$ whose colors are
within $\pm 0.2$ magnitudes of the
color-magnitude relation. This surface density profile, extended by a power
law fit, gives $R_{30} = 59
\arcsec$. In the following, we take $R_{30}=0.5$\,\h50min\,Mpc
(equivalent to $53\arcsec$). This number is very similar to values
measured by Butcher \& Oemler (1984) for clusters at lower redshift.

We correct our spectroscopic sample for incompleteness
by assigning each galaxy a weight equal to the inverse of the
magnitude selection function
(cf.\ Sect.\ \ref{complete.sec}). The corrected number of
cluster galaxies with $I_{F814W}\leq 22.5$
and $R<R_{30}$ is 67. Eleven (16\,\%) are blue according
to the Butcher \& Oemler criterion, where we have converted
restframe $U-B$ to $B-V$ using $\Delta (U-B) = 1.4 \Delta (B-V)$,
as derived from models by Worthey (1994).

Our spectroscopic sample extends to
$I_{F814W} \approx 22.5$.
The magnitude limit that is equivalent to the Butcher \& Oemler
limit of $M_V = -20$ in the restframe is $I_{F814W} \sim 24$,
where we have used
Eq.\ \ref{trafomag.eq}, and $B-V = 1$ in the restframe.
To determine the number of blue galaxies with $22.5 < I_{F814W}
<24$ we cannot use our sample of confirmed members,
and hence we used the full photometric dataset, and applied
a correction for field contamination.
We predict the number of field galaxies in this magnitude
range by fitting a power law to the differential galaxy counts
at $20<I<22.5$. These counts were determined from our sample
of confirmed field galaxies, and corrected for incompleteness.
The power law fit has exponent $\alpha_I = 0.33$, similar to
values determined from deep photometric surveys (e.g., Smail et al.\
1995). Of 115 galaxies with $22.5 < I_{F814W} < 24$ and $R<R_{30}$,
$27$ are predicted to be
field galaxies. Assuming the color distribution
of the field galaxies is identical to that of field galaxies
in our spectroscopic sample, we find $23$ of $88$
cluster galaxies with $22.5 < I_{F814W} < 24$  and $R<R_{30}$ are blue.

Combining these numbers with those determined from the
spectroscopic sample, we find a Butcher \& Oemler blue fraction
$f_B = 0.22$, with a random error of $\pm 0.04$. From experimenting
with the value of $R_{30}$, the completeness corrections,
and the field correction we estimate the systematic
uncertainty is $\pm 0.03$. Assuming the random error and the
systematic error can be added in quadrature, we find
$f_B = 0.22 \pm 0.05$. 

Figure \ref{bo.plot} shows the redshift evolution of the fraction of
blue galaxies in rich clusters.
The blue fraction in \1054{} is similar to typical blue
fractions in optically selected clusters at $z \sim 0.4$,
and higher than typical blue fractions in X-ray selected clusters
at $z \sim 0.2$. The extremely high
fractions of blue galaxies found by Rakos \& Schombert (1995) in
clusters at $z>0.5$ ($f_B \sim 0.8$ at $z \sim 0.8$) are not confirmed
here. This may in large part be due to differences in procedure:
Rakos \& Schombert (1995) did not impose a
strict radial limit, and determined blue fractions by comparing the
colors of galaxies in their high redshift clusters to those of
present-day elliptical galaxies. This method does not take passive
evolution of the ridge-line of the CM relation into account.


\vbox{
\begin{center}
\leavevmode
\hbox{%
\epsfxsize=8cm
\epsffile{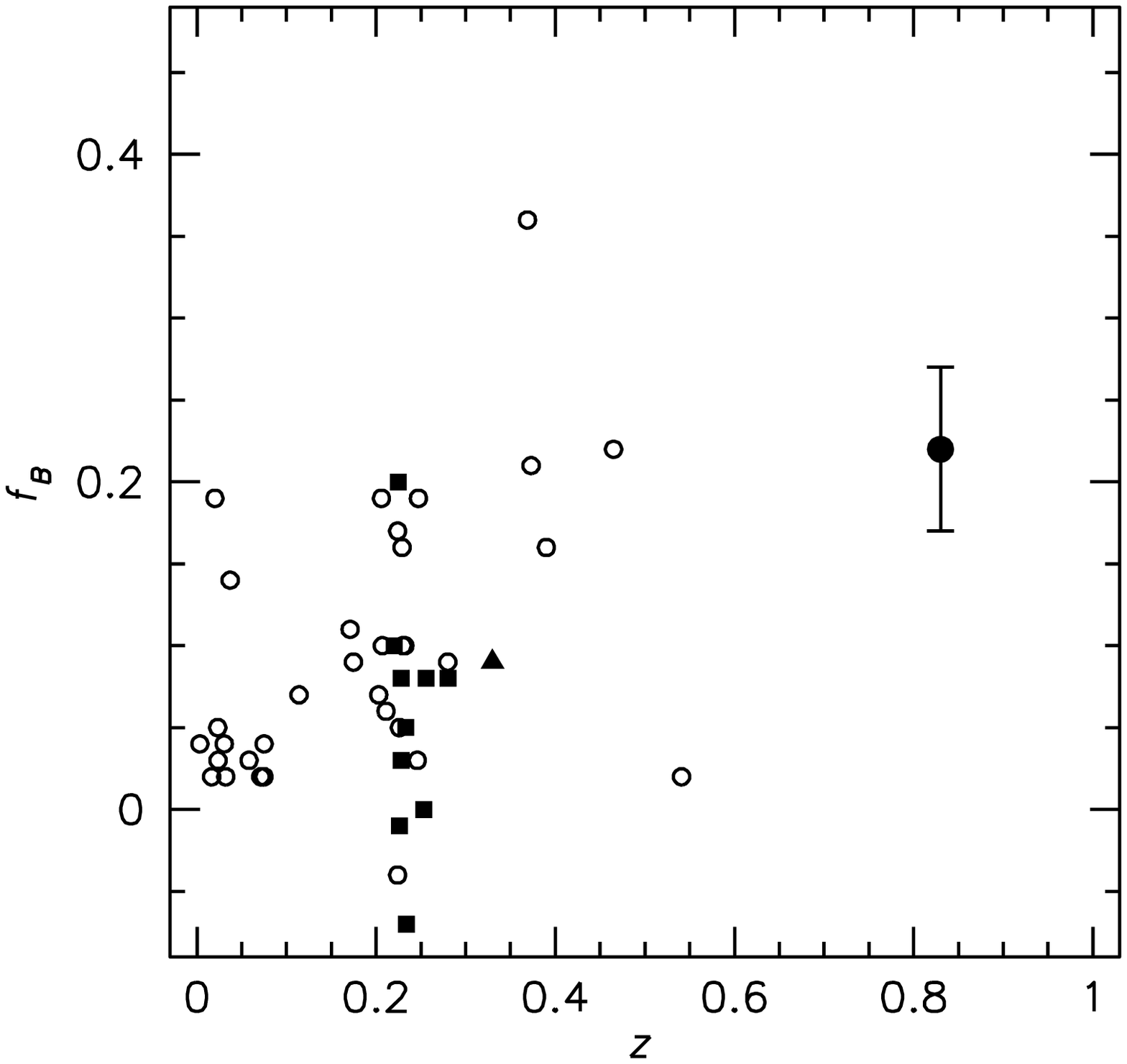}}
\begin{small}
\figcaption{\small
The evolution of the Butcher-Oemler blue fraction
with redshift. The large symbol at $z=0.83$ is \1054{}.
Circles are clusters
from Butcher \& Oemler (1984), squares are clusters
from Smail et al.\ (1998),
and the triangle is CL\,1358+62 from Fabricant et al.\ (1991). For clarity
no errorbars are shown on the datapoints from the literature;
typical errors are $\sim 0.03$ for low redshift
clusters and $\sim 0.08$ for intermediate redshift clusters. Filled
symbols are X-ray selected clusters. The blue fraction in
\1054{} does not exceed typical values for clusters at $z \sim 0.4$.
\label{bo.plot}}
\end{small}
\end{center}}

We note that three of the four spectroscopically confirmed blue
galaxies within $R_{30}$ (galaxies 1422, 1459 and 1650) are late-type
spiral galaxies; the eight other spiral galaxies with $R<R_{30}$ are
red.  One blue galaxy (1035) is classified as merger.
Spectroscopically, 1035 and 1650 are E+A galaxies, galaxy
1459 has strong H$\delta$ absorption and weak [O\,{\sc ii}]
3727\,\AA{} emission, and galaxy 1422 has an early-type spectrum.

\section{Evolution of the Color-Magnitude Relation of
Early-Type Galaxies from $z=0$ to $z=1$}
\label{evoscat.sec}

\subsection{Evolution of the Slope}

The slope and scatter of the CM relation of early-type galaxies can be
compared to data at lower redshift. Fig.\ \ref{slopeevo.plot} shows
the evolution of the slope of the CM relation as a function of
redshift. Data points are from Bower et al.\ (1992)
for the Coma cluster, van Dokkum
et al.\ (1998a) for CL\,1358+62 at $z=0.33$, and Ellis et al.\ (1997)
for three clusters at $z \approx 0.55$.  We have used the Worthey
(1994) models to transform $U-V$ (Bower et al.\ (1992)
and Ellis et al.\ 1997) and
$B-V$ colors (van Dokkum et al.\ 1998a) to $U-B$ colors. For solar
metallicity these models give $\Delta (U-B) = 1.4 \Delta (B-V)$ and
$\Delta (U-B) = 0.6 \Delta (U-V)$.

The samples of Bower et al.\
(1992) and Ellis et
al.\ (1997) only cover the inner parts of the clusters. Furthermore,
in these studies
colors were measured inside fixed apertures of 10\,\h50min\,kpc
rather than within effective radii. For
CL\,1358+62 and \1054{} we determined the slope for the inner parts of
the cluster as well as for the whole sample. To enable a direct
comparison to Bower et al.\ (1992) and Ellis et al.\ (1997) we
also determined the slope of the CM relation in the inner
parts of the cluster with color measurements in
a fixed aperture of 10\,\h50min\,kpc.


\vbox{
\begin{center}
\leavevmode
\hbox{%
\epsfxsize=8cm
\epsffile{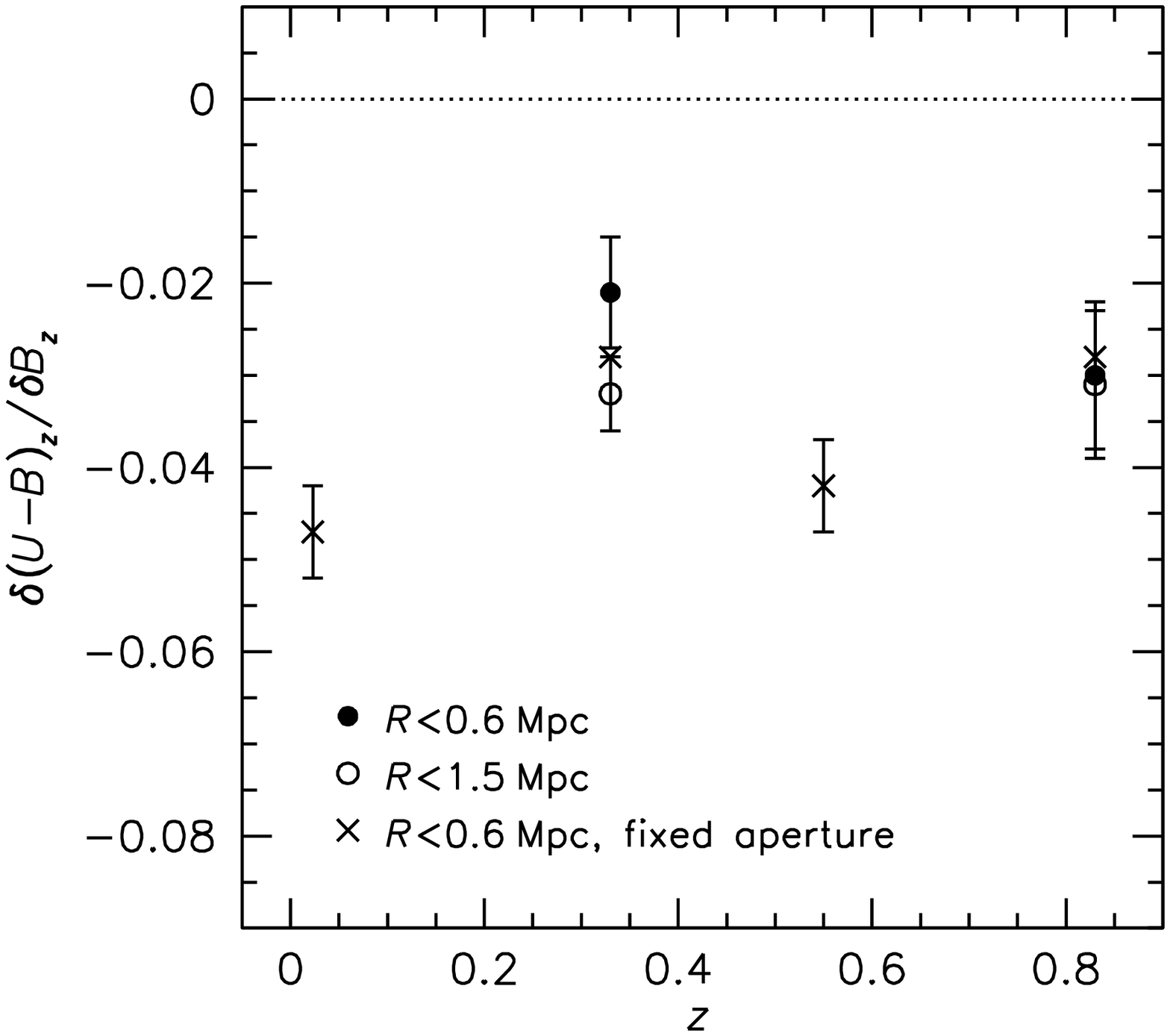}}
\begin{small}
\figcaption{\small
Evolution of the slope of the CM relation of early-type galaxies,
based on HST photometry. In order of increasing redshift the data
points are from Bower et al.\ ($z=0.023$), van Dokkum et al.\ (1998a)
($z=0.33$), Ellis et al.\ (1997) ($z=0.55$), and this study
($z=0.83$).  The data are consistent with a constant slope with
redshift.
\label{slopeevo.plot}}
\end{small}
\end{center}}

A linear fit to the evolution of the slope in the inner parts of the clusters
gives $\delta (U-B)_z/\delta B_z = (0.013 \pm 0.016) z - (0.041 \pm
0.008)$. We conclude that there is no evidence that
the slope of the CM relation depends on redshift.
If the slope in the relation at $z=0$ is (partly) due to
a systematic age gradient, such that high luminosity galaxies have
higher luminosity weighted ages than low luminosity galaxies, one might
expect the slope to become steeper at higher redshift (e.g., Kodama et
al.\ 1998). We see no such effect in the data
presented here. This result is in agreement with
the studies of Stanford et al.\ (1998) and
Kodama et al.\ (1998), who used ground based photometry and HST
morphologies to determine the
slope and scatter of the CM relation of early-type galaxies
in clusters at $0 < z < 1$.

\subsection{Evolution of the Scatter: the Progenitor Bias}

The evolution of the scatter in the CM relation of early-type galaxies
is shown in Fig.\ \ref{scatevo1.plot}. Solid symbols are from
the same sources as in Fig.\ \ref{slopeevo.plot}, and
are based on WFPC2 colors (except for the Coma cluster at $z=0.02$). 
Open symbols are taken from Stanford et al.\ (1998), and are
based on ground based photometry. Stanford et al.\ (1998) assumed
their ``blue -- red'' colors are close to restframe
$U-V$, and we transformed their measurements using
$\Delta (U-B) = 0.6 \Delta$(``blue -- red'').


The scatter in the CM relation of early-type galaxies is low at all
redshifts, consistent with the results of Stanford et al.\ (1998). In
particular, the scatter at $z=0.83$ is almost identical to the scatter
at $z=0$. This may seem remarkable, because the scatter in the CM
relation might be expected to increase with redshift due to increasing
fractional age differences between galaxies (see, e.g., Bower et al.\
1992, van Dokkum et al.\ 1998a, Ferreras \& Silk 2000).  The
underlying assumption of this argument is that early-type galaxies are
drawn from the same sample at high and low redshift, and that they
evolve in a regular way between $z=1$ and $z=0$. However, the low
early-type fraction and the presence of the mergers demonstrate that
the sample of early-type galaxies in \1054{} only forms a subset of
the population of present-day early-type galaxies.

\vbox{
\begin{center}
\leavevmode
\hbox{%
\epsfxsize=8cm
\epsffile{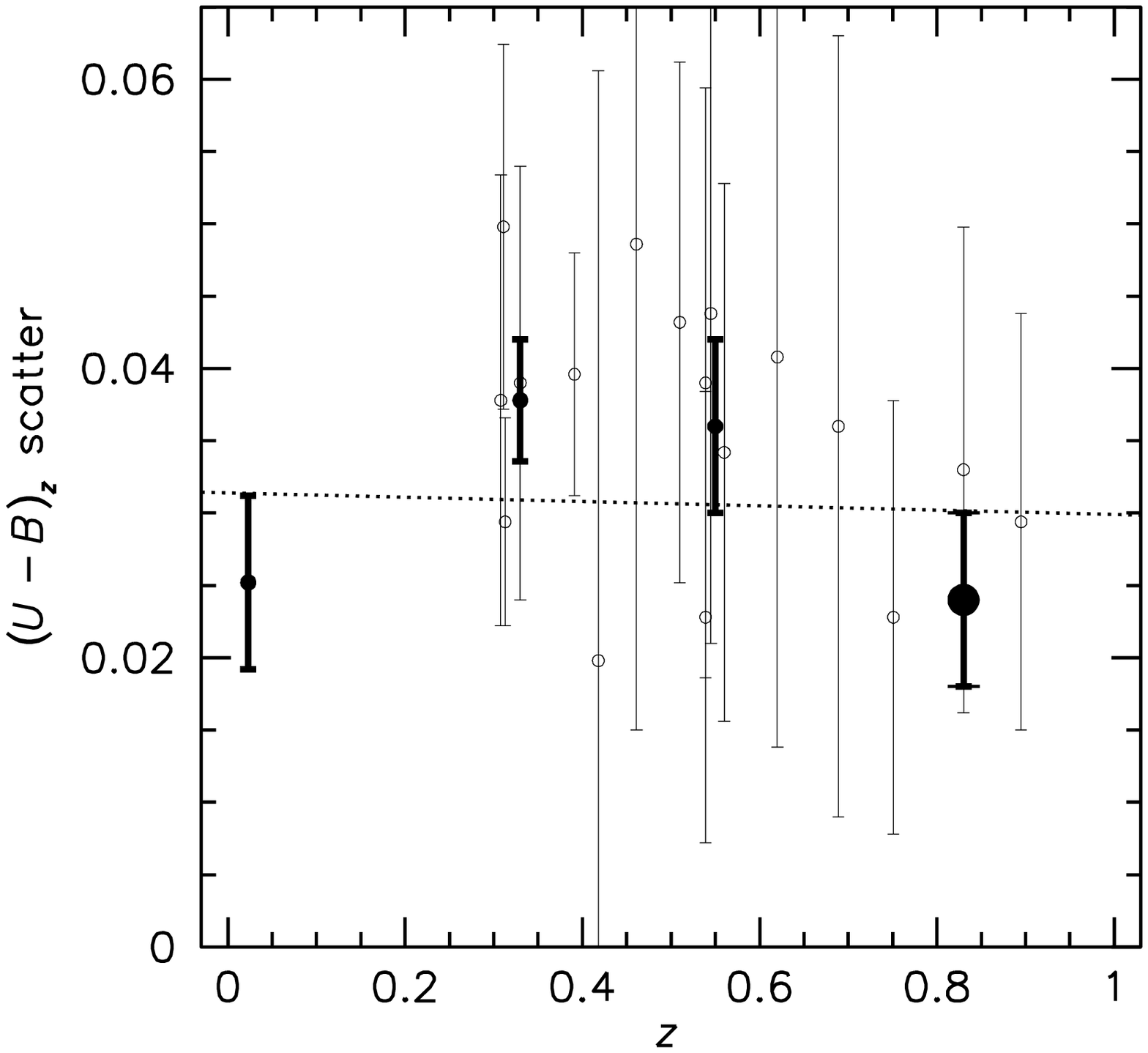}}
\begin{small}
\figcaption{\small
Evolution of the scatter in
the CM relation of early-type galaxies. Filled data points are
from the same sources as those in Fig.\ \ref{slopeevo.plot},
and are based on HST photometry (except the Coma cluster).
Open data points are from Stanford et al.\ (1998) and are based
on ground based colors. Note that the HST
measurements have smaller uncertainties than the ground based
measurements. The broken line is a linear least squares fit
to the HST photometery. The data are consistent with a constant
scatter with redshift.
\label{scatevo1.plot}}
\end{small}
\end{center}}

Figure \ref{scatevo2.plot} shows the scatter in the CM relation of
{\em all} known progenitors of present-day early-type
galaxies. Specifically, this includes the mergers in \1054{} since
they will probably evolve into early-type galaxies.  The intrinsic
scatter in
the combined sample of mergers and early-type galaxies in \1054{}
is twice as high as the scatter in the early-type galaxies alone, and the
data indicate an increase in the scatter by a factor two between $z=0$
and $z=1$. We fitted linear relations to the data shown in Fig.\
\ref{scatevo1.plot} and Fig.\ \ref{scatevo2.plot}.  These fits are
indicated by broken lines.  The best fitting relation changes from a
constant to substantial evolution.


We note that in addition to the mergers
a fraction of the spiral galaxies in \1054{} may also evolve
into early-type galaxies (e.g., Dressler et al.\ 1997). It is
very hard to establish which, if any, of the spiral galaxies in
\1054{} are destined to become S0 galaxies.
If the mergers and all red spiral
galaxies (i.e., those within 0.2 magnitudes of the CM relation)
evolve into early-type galaxies, the scatter in the color-magnitude
relation of all progenitors of present-day early-type galaxies is
$\approx 0.063$ at $z=0.83$.

These results show that the properties of samples of
early-type galaxies in distant clusters are different from the properties
of the full sample of progenitors of present-day early-type
galaxies.
Distant early-type galaxies are the progenitors of the oldest present-day
early-type galaxies. This ``progenitor bias'' can lead to biased age
estimates. As demonstrated here, the
scatter in the color-magnitude
relation is particularly sensitive to this bias.
In a forthcoming paper we model the
redshift evolution of the scatter in the CM relation, taking
morphological evolution into account.

An open question is whether the galaxy
population in \1054{} is typical for its redshift.  The cluster is
young and still forming, and the high merger fraction could be related
to this process (van Dokkum et al.\ 1999).  If this is the case, this
epoch of enhanced merging could occur at different redshifts for
different clusters.

\vbox{
\begin{center}
\leavevmode
\hbox{%
\epsfxsize=8cm
\epsffile{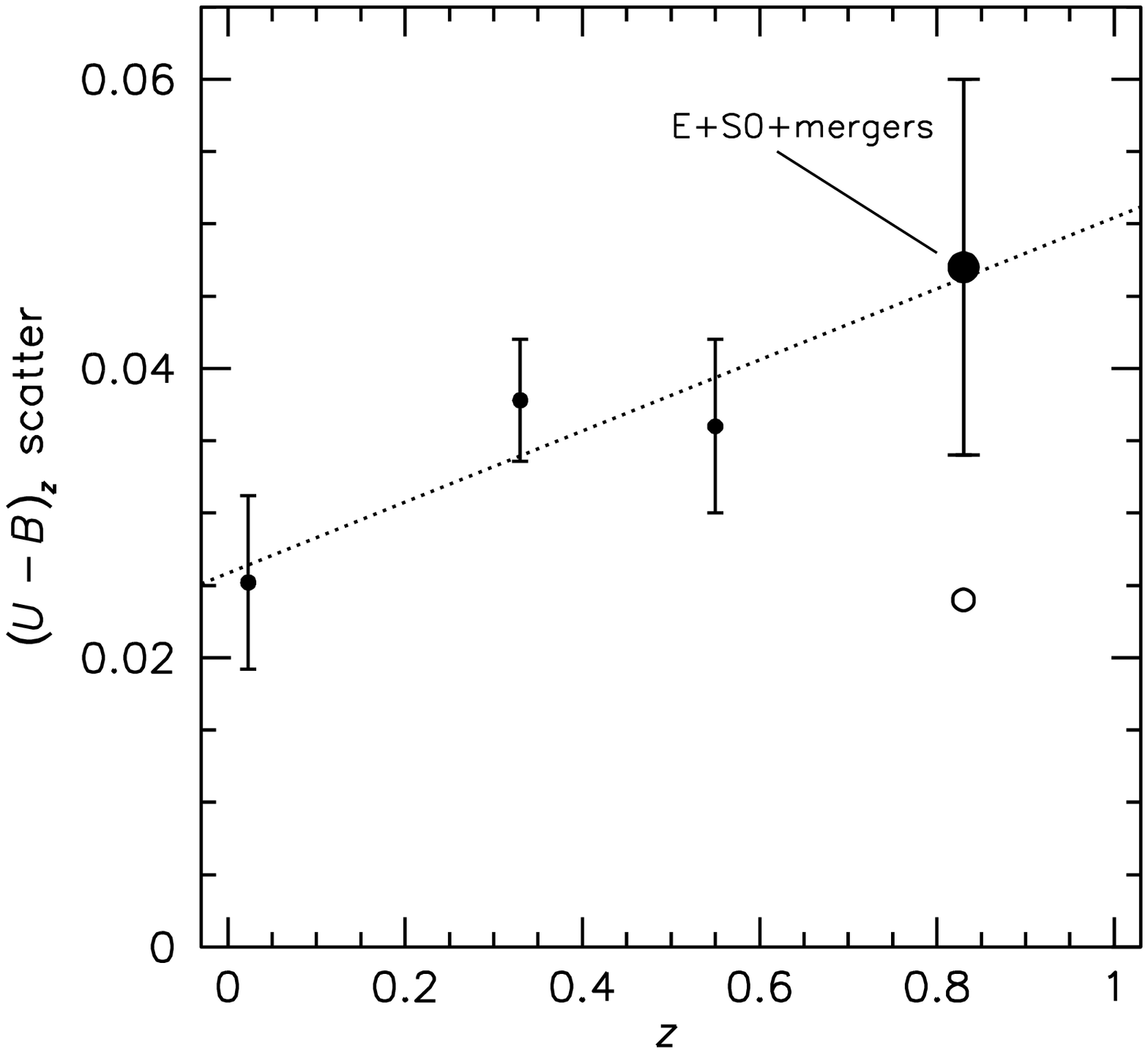}}
\begin{small}
\figcaption{\small
Evolution of the scatter in the CM relation
of all progenitors of present-day
early-type galaxies. This includes the mergers in \1054{} at $z=0.83$.
For comparison, the open symbol at $z=0.83$ shows the
scatter of the early-type galaxies alone.
The points at lower redshift are identical to those in
Fig.\ \ref{scatevo1.plot}.
The broken line is a linear least squares fit to the data.
The scatter in the full sample of progenitors of present-day
early-type galaxies increases by a factor $\sim 2$ between $z=0$ and
$z=1$.
\label{scatevo2.plot}}
\end{small}
\end{center}}

\section{Summary and Conclusions}

We have presented a study of the galaxy population in
the rich cluster \1054{} at $z=0.83$.
Our sample of 81 $I$-band selected galaxies is
currently the largest sample of confirmed cluster galaxies at $z>0.5$
observed with HST.  One of the most striking results of our survey of \1054{}
is the high fraction of mergers among the $L>L_*$ cluster population
(van Dokkum et al.\ 1999). In the present paper we focussed on the
early-type fraction, the Butcher-Oemler effect, and the
color-magnitude relation in \1054.

The early-type fraction in \1054{} is much lower than that in
rich clusters at intermediate and low redshift.
Combining our data with data from the literature
we find a continuous decrease in the early-type fraction with
redshift. The early-type fraction changes by a factor $\sim 2$
from $z=0$ to $z=0.83$.
This result is robust and independent of the classification
procedure. It is reasonable to assume that the mergers in \1054{}
will evolve into early-type galaxies.
This implies that the early-type fraction
in this cluster will increase from $44$\,\% to
61\,\% in approximately 1\,Gyr.

\1054{} displays a similar Butcher-Oemler effect as clusters at
intermediate redshift: $f_B = 0.22 \pm 0.05$.
This may appear to be at variance with the low early-type fraction,
but is a result of the fact that most spiral galaxies are red, similar to
spiral galaxies in nearby clusters (e.g., Butcher \& Oemler 1984).
Of eleven spiral galaxies within $R=R_{30}$, only three are blue.
The blue fraction in \1054{} is much lower than values reported
by Rakos \& Schombert (1995) for clusters
at $z>0.6$. This is probably
due to the fact that these authors
used a different procedure to measure the blue fraction.

We studied the CM relation as a
function of morphology. The early-type galaxies follow a remarkably
tight and well defined relation, with an
observed scatter of $0.029 \pm 0.005$
in restframe $U-B$. There is no significant dependence of the CM relation of
early-type galaxies on radius in the cluster; our small sample
precludes us from confirming or ruling out a similar trend as found
for S0 galaxies in the cluster CL\,1358+62 at $z=0.33$
(van Dokkum et al.\ 1998a).
The spiral galaxies exhibit a large range in their colors; they are offset
to the blue by $0.22$ magnitudes with respect to the CM relation
of the early-type galaxies, but many are as
red as early-type galaxies of the same luminosity.
This could be due to a spread in age, metallicity, and/or dust content.

The mergers are only
slightly bluer than the early-type
galaxies, by $0.07 \pm 0.02$ magnitudes. The
scatter in their CM relation is $0.07 \pm 0.04$. Only three
of the thirteen mergers are more than 0.2 magnitudes bluer than
the CM relation; these are the only mergers with detected
[O\,{\sc ii}] 3727\,\AA{} emission.
It may appear surprising that the mergers are so red; mergers in the
nearby Universe are typically blue
and have high star formation rates (Liu \& Kennicutt 1995).
However, most of the spiral galaxies in \1054{} are red as well, and
the red colors of the mergers
are probably a result of the environment:
galaxies in and near a rich cluster are likely redder
than typical field galaxies. A quantitative demonstration of this
effect is that the median color of the mergers ($\langle U-B \rangle_z
= 0.36$) is very similar to the median color of the full
sample of cluster galaxies ($\langle U-B \rangle_z = 0.41$).

The slope and scatter of the CM relation of the early-type galaxies
in \1054{} are very
similar to those of clusters at lower redshift, confirming previous
studies. However, the presence of the mergers and the low
early-type fraction in \1054{} demonstrate that
early-type galaxies in distant clusters are a biased set, and
not representative of the full sample of progenitors of present-day
early-type galaxies. This ``progenitor bias'' has a large effect
on the scatter in the color-magnitude relation: the scatter in
the full sample of all progenitors of present-day early-type
galaxies (i.e., including the mergers in \1054{}) shows strong
evolution with redshift. The scatter increases by a factor $\sim 2$
between $z=0$ and $z=1$.
The main uncertainty in this result is the assumption that
the galaxy population in \1054{} is typical for its
redshift.

It is clearly important to extend this type of analysis to other
clusters, and to higher redshift.  Of particular interest  is
the evolution of the early-type fraction, since it may be
a fairly robust measure of morphological evolution.
The large format of the HST imaging
was very important in our study, because the mergers typically reside
at large radii (van Dokkum et al.\ 1999), and it allowed efficient
spectroscopic follow-up. This study demonstrates that
the combination of extensive spectroscopy
and large format, high resolution imaging can provide important
insights into the assembly and evolution of cluster galaxies.

\begin{acknowledgements}
We thank the anonymous referee for constructive and valuable comments,
which improved the paper significantly. We thank G. Luppino
for providing us with a ground based $R$ band image of \1054.
We appreciate the help of the staff of the W. M. Keck
Observatory. Support from STScI grant
GO07372.01-96A is gratefully acknowledged.
\end{acknowledgements}

\begin{appendix}

\section{Details of the Spectroscopic Survey}

\subsection{Reduction}
\label{specred.sec}

The reduction involved bias subtraction, flatfielding, removal of
cosmic rays, rectification, and sky subtraction. Critical steps in the
reduction process were flatfielding and cosmic ray removal.
Flatfielding is difficult because the Tektronix CCD has severe
fringing at $\lambda \gtrsim 7500$\,\AA.  The fringes have characteristic
scales of
$\lambda \sim 20$\,\AA{}, and amplitudes to $\sim 8$\,\%. The
signal-to-noise in the sky subtracted spectra critically depends on
the success of removing the fringe pattern. For this reason we took
flatfield exposures at a range of hour angles and grating angle
settings, bracketing the hour angle and grating angle setting at the
time of the observations. The flatfields that produced the lowest
noise in the sky lines were used. This procedure reduced the amplitude
of the fringes to $\approx 0.5$\,\%.

Cosmic rays were removed as described by van Dokkum
\& Franx (1996). The galaxy spectra and sky lines were fitted with
appropriate functions and subtracted. Pixels that deviated
significantly from the expected noise in the subtracted functions were
flagged as possible cosmic rays. Several slits in the masks were
tilted in order to cover more than one object with the same slit.
Such tilted slits produce tilted sky lines in the 2D spectra, and
these were not adequately removed by the fitting functions. Residuals
from sky lines in these slits were removed interactively.  After
subtraction of the cosmic rays, the frames were carefully checked by
eye to ensure that only cosmic rays were removed.

For each row in each
slitlet a wavelength solution was obtained from the sky lines.
Wavelengths of
unblended lines (the [O\,{\sc i}] 5577, 6300\,\AA{} lines
and the OH P1 lines) were taken from the Osterbrock et al.\ (1996) catalog.
The 2D galaxy spectra were corrected for the S-distortion and
transformed to a common wavelength scale. Sky spectra were
determined from the edges of each slitlet, and subtracted. Finally, 1D
spectra were extracted from the 2D spectra using {\sc apsum} in IRAF.

\subsection{Determination of Spectral Types}
\label{lines.sec}

We have measured equivalent widths of the [O\,{\sc ii}] 3727\,\AA{}
emission line and the H$\delta$ 4102\,\AA{} Balmer line. The strength
of the [O\,{\sc ii}] line is a measure of ongoing star formation, and
the strength of the H$\delta$ line is sensitive to recent star
formation activity (e.g., Barbaro \& Poggianti 1997).  We used the line
strength index definitions of Fisher et al.\ (1998).  Line strengths
are positive if the line is in absorption, and negative if in
emission.  The restframe equivalent widths and their associated errors
are listed in Table 1. For most galaxies the errors are a few
Angstroms.

Spectral types were assigned to the galaxies
based on the strengths of [O\,{\sc ii}] and H$\delta$.
The definitions of the spectral types are also
analogous to those of Fisher et al.\ (1998). The spectral types are
``absorption'', ``emission'', and ``E+A'' (Dressler \& Gunn 1983).
Emission line galaxies have
[O\,{\sc ii}] 3727\,\AA{} emission line strength greater than
5\,\AA{}. Galaxies with [O\,{\sc ii}] emission
line strength
less than 5\,\AA{} and Balmer absorption line strength greater than
4\,\AA{} are classified as ``E+A'' galaxies. Galaxies lacking significant
[O\,{\sc ii}] emission and strong H$\delta$ absorption have spectral
type ``absorption''. The spectral types are listed in Table 1.
Spectral types which within $1\sigma$ of another
type (as determined from the errors in [O\,{\sc ii}] and
H$\delta$) are labeled with a question mark.

\section{Reduction of WFPC2 Data and Photometry}

\subsection{Observations and Initial Reduction}

The \1054{} field was observed with the F606W and F814W filters on
1998 May 30.  We obtained a mosaic of six slightly overlapping
pointings. Six exposures were taken in each filter and at each
position, giving a total of 72 exposures.  In order to improve the
sampling the six exposures at each position were split in two sets of
three, with relative offsets between the sets of 5.5 pixels in $x$ and
$y$.  Furthermore, the three exposures in each set were shifted
relative to each other by $\pm 3.0$ pixels, to facilitate the
identification of hot pixels.  Total exposure times were 6500 s in
each filter for each pointing. The layout of the field is shown in
Fig.\ \ref{colfield.plot}.

The pipeline reduction was performed at the Space Telescope Science
Institute.  Further processing involved masking of bad pixels and bad
columns, shifting the exposures to a common position on the sky, sky
subtraction, identifying cosmic rays and hot pixels, and combining the
images.  The most labour intensive step in this process is the cosmic
ray removal. We experimented with the {\sc crreject} task in the {\sc
stsdas} package, but found that the quality of the combined image is
quite sensitive to the choice of input parameters. In particular, the
noise in the final combined image can be higher than expected from the
noise in the individual input images.  We therefore followed a
different strategy. We compared the exposures with integer pixel
shifts in each set of three, and removed cosmic rays in several steps.
First, the expected noise $\sigma_{\rm exp}$ in each pixel was
calculated from the minimum of the three exposures. Then, the task
{\sc gcombine} in IRAF was used to identify cosmic rays, with
$\sigma_{\rm exp}$ as the input noise model.  The output of {\sc
gcombine} is a cosmic ray cleaned average of the three exposures, an
image containing the number of cosmic rays found, and an output noise
map $\sigma_{\rm true}$ based on the actual variation in each pixel.
Finally, the noise model and the measured noise were compared to
identify pixels affected by cosmic rays in two of the three
exposures. The method proved to work very well. In particular, in no
cases were the central pixels of stars or galaxies mistaken for cosmic
rays.  After identifying the cosmic rays each set of three exposures
with integer pixel shifts was averaged, excluding the flagged pixels.
As a result there are two reduced images of each field and in each
filter, shifted with respect to each other by 5.5 pixels.

\subsection{Combination of Interlaced Images}
\label{interlace.sec}


As is well known, the $0\farcs 1 \times 0\farcs 1$ WFC pixels
undersample the HST WFPC2 PSF.  The sampling can be improved by
obtaining multiple exposures of the same field, offset by subpixel
shifts. The shifted exposures can be ``drizzled'' on a finer grid (see
Fruchter \& Hook 1999 and references therein).  If the offsets between
the exposures are precisely 0.5 pixels, the images are interlaced. In
this special case there is a one-to-one correspondence between pixels
in the original image and in the higher resolution output image. The
improvement in the sampling is a factor $\sqrt{N}$, with $N=2$ or
$N=4$ the number of independent positions.

\begin{figure*}[t]
\epsfxsize=7cm
\epsffile{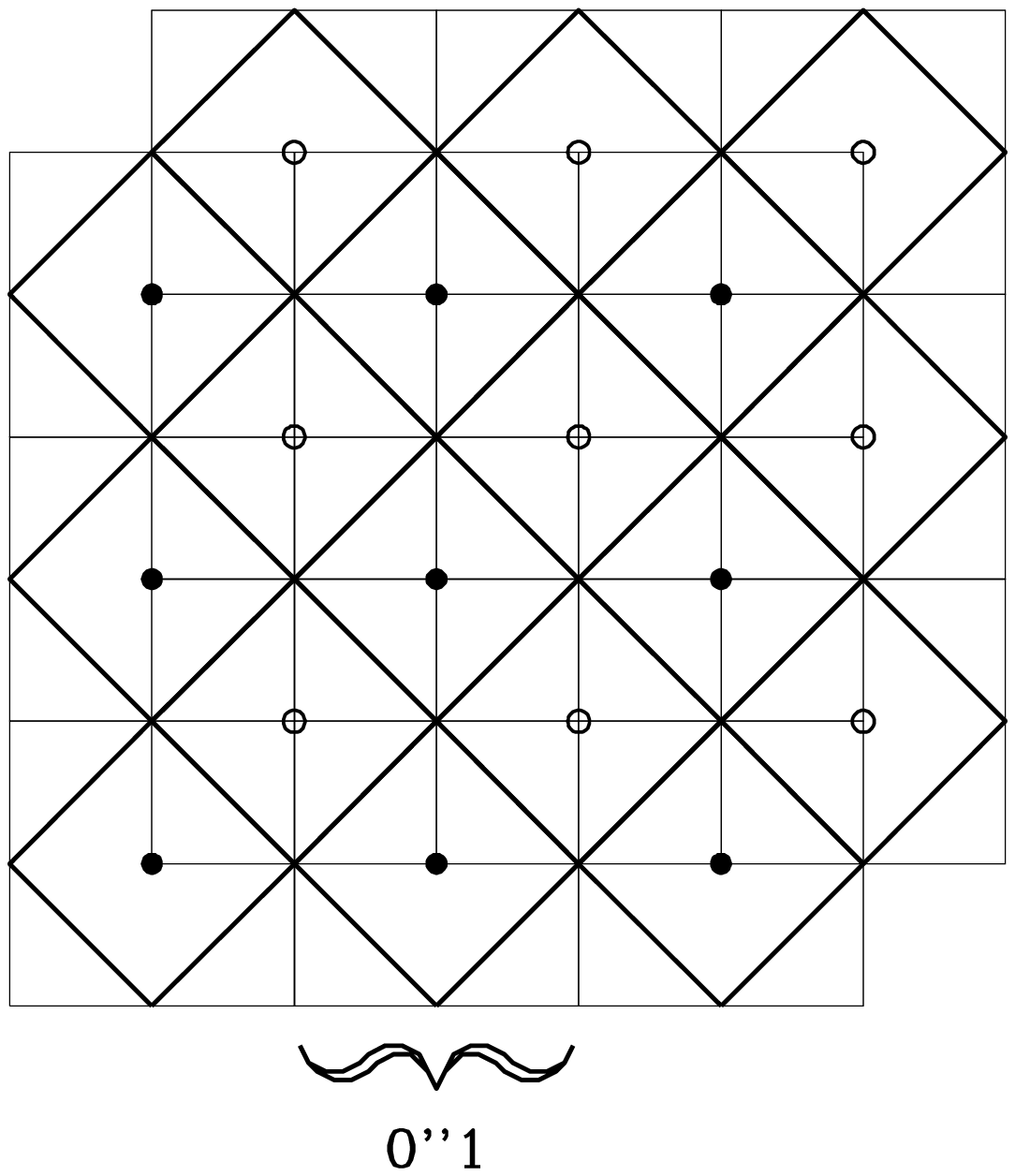}
\caption{\small
Illustration of interlacing using two independent positions. Solid
circles are the centers of pixels in the first exposure, open circles
are the centers of pixels in the second, offset, exposure. The solid
circles and open circles sample a grid that is rotated by
$45^{\circ}$ with respect to the original grid. The output pixels are
indicated by heavy lines. The output pixel size is $0\farcs1/\sqrt{2}$.
\label{sampling.plot}}
\end{figure*}

The present observations are taken at two independent positions,
offset by $\pm 5.5$ pixels. As illustrated in Fig.\
\ref{sampling.plot}, these two exposures sample an $(x,y)$ grid which
is rotated by $45^{\circ}$ with respect to the original grid.  We
combined the two offset exposures of each field by copying the pixels
on a finer grid.  The transformations from the offset input images
${I}^A$ and ${I}^B$ to the interlaced output image ${I}^{AB}$ are
\begin{eqnarray}
{I}^{AB}_{n+i-j,i+j-1} & =& {I}^A_{i,j}\\
{I}^{AB}_{n+i-j,i+j} & =& {I}^B_{i,j},
\end{eqnarray}
with $n \times n$ the size of the input images, and $i,j =
1,\,2,\,...,\,n$. The output image has a size $2n \times 2n$.  The
area of the output image containing data is $\sqrt{2}n \times
\sqrt{2}n = 2n^2$ pixels.  The pixel size is $0\farcs1/\sqrt{2} =
0\farcs071$.


\subsection{Image Restoration}
\label{imrest.sec}

\begin{figure*}
\epsfxsize=17cm
\epsffile{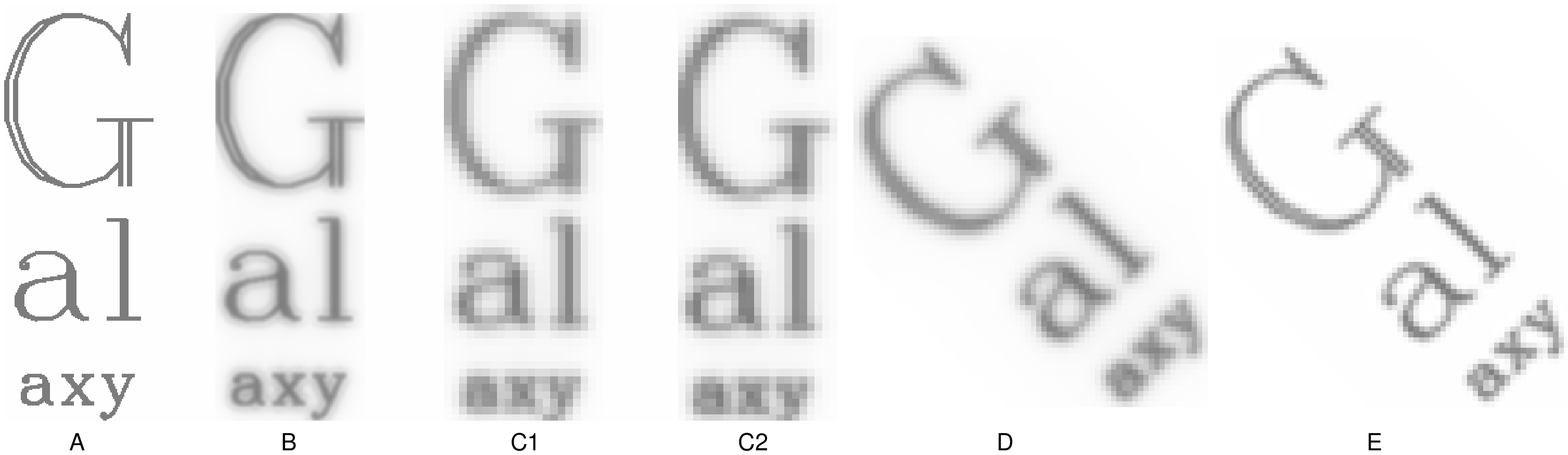}
\caption{\small
Effect of resolution enhancement as a result of
interlacing and subsequent {\sc clean}ing.
{\bf A}: Original (``true'') image of dimensions
$2\farcs 1 \times 5\farcs 5$; {\bf B}:
image A convolved with HST WFPC2 PSF;
{\bf C1}, {\bf C2}: image B sampled at WFPC2 $0\farcs 1$
resolution and convolved with the subpixel response function, at two
positions with relative offsets of $0 \farcs 05$ in $x$ and $y$.
Note that most of the degradation of image A is a result of the
coarse sampling.
{\bf D}: Interlaced combination of images C1 and C2, at
$0\farcs 071$ resolution; {\bf E}: {\sc clean} restoration
of image D.
\label{galaxy.plot}}
\end{figure*}

Although the interlacing improved the sampling by a factor $\sqrt{2}$,
the improvement in resolution is smaller. This is caused by the
wings of the PSF, and the subpixel response function (e.g.,
Krist 1995). In order to
correct the photometry for these effects
the interlaced images were restored using
{\sc clean} (H\o{}gbom 1974).
The {\sc clean} algorithm requires a model for the PSF.
For all cluster galaxies subsampled PSFs appropriate for the positions of the
galaxies on the WFC chips were created using Tiny Tim (Krist 1995),
version 4b.
The exposures were mimiced by creating a copy of each subsampled PSF and
shifting it by $0\farcs05$ in $x$ and $y$.
The PSFs were then rebinned to the WFC
resolution, and convolved with the subpixel response function using the
kernel given in Krist (1995). Finally, each PSF was combined with its
$0\farcs05$ shifted copy in the same way as the actual exposures.
The effect of interlacing and subsequent {\sc clean}ing on noiseless
data is demonstrated in Fig.\ \ref{galaxy.plot}. Note that most of the
degradation of image {\bf A} is caused by the coarse sampling of the PSF by
the $0\farcs 1$ pixels. 


\subsection{Photometry}

\begin{figure*}[p]
\epsfxsize=17cm
\epsffile{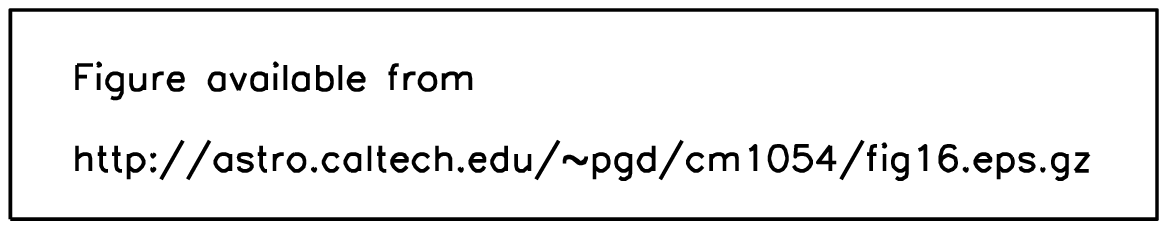}
\caption{\small
\label{morph_mos.plot}
Color images of the 42 most luminous
spectroscopically confirmed cluster members in
\1054{}. The images are created from the F606W and F814W exposures,
and are sorted by total F814W magnitude.
Each image is $5\farcs9 \times 5\farcs9$;
at the distance of \1054, this corresponds to $63 \times
63$\,\h50min\,kpc. Note the large fraction of mergers among
the most luminous galaxies in the cluster, and their fairly red
colors.}
\end{figure*}

Figure \ref{morph_mos.plot} shows color images of the 81
spectroscopically confirmed cluster members in the HST WFPC2 mosaic.
Total (``best'') magnitudes of these galaxies were determined from the
HST data using the SExtractor program (Bertin \& Arnouts 1996). Colors
were determined from the HST data using {\sc phot} in {\sc iraf}.  The
colors were measured within the effective radii of the galaxies, which
were derived from fits of the galaxy images to $r^{1/4}$ laws
convolved with the PSF (see van Dokkum et al.\ 1998a).  As explained
in Sect.\ \ref{imrest.sec} we used the {\sc clean}ed images for these
measurements.


The observed F606W and F814W bands roughly correspond to restframe
$U$ and $B$ for objects redshifted to $z=0.83$.
Following the method outlined in
van Dokkum \& Franx (1996) transformations were computed
from the observed bands to redshifted $B$ (denoted $B_z$),
and $(U-B)_z$ colors:
\begin{equation}
\label{trafomag.eq}
B_z = {\rm F814W} - 0.03 ({\rm F606W} - {\rm F814W}) + 1.23
\end{equation}
\begin{equation}
(U-B)_z = 0.76 ({\rm F606W} - {\rm F814W}) - 1.14.
\end{equation}
The zero points of the F814W and F606W filters are taken from the
HST Data Handbook (STScI, Baltimore).
Total $B_z$ magnitudes were
converted to total absolute $M_B^{\rm T}$ magnitudes using
$H_0 = 50$\,\kms\,Mpc$^{-1}$ and $q_0 = 0.15$.

The photometric accuracy can be assessed empirically, by comparing
the two independent measurements of colors of objects in
the overlapping regions between the six HST pointings.
Note that this procedure limits the effects resulting from uncertainties
in the CTE degradation of the WFPC2 chips.
From this comparison
we find that the average photometric uncertainty in the color
is $0.016$ for a single measurement.
In our analysis of the scatter in the CM relation, this number
is subtracted in quadrature from the measured
scatter.

\end{appendix}

\newpage

\begin{table}
\begin{center}
{ {\sc TABLE 3} \\
\sc Offset and Scatter of the CM Relation} \\
\vspace{0.1cm}
\begin{tabular}{lrrccc}
\hline
\hline
Sample & $N$ & Offset & Error & Obs.\ Scatter & Error\\
\hline
E+S0 & 30 & $0.000$ & 0.005 & 0.029 & 0.005\\
S & 26 & $-0.219$ & 0.047 & 0.255 & 0.025\\
M/P & 13 & $-0.066$ & 0.019 & 0.074 & 0.040\\
E+S0+M/P & 43 & $-0.008$& 0.007 & 0.050 & 0.013 \\
All & 81 & $-0.018$ & 0.008 & 0.081 & 0.029\\
\hline
\end{tabular}
\end{center}
\end{table}

\begin{small}
\begin{table}[p]
\begin{center}
{ {\sc TABLE 1} \\
\sc Spectroscopic Data} \\
\vspace{0.1cm}
\begin{tabular}{rcrrrrcrcrrrrc}
\hline
\hline
Id &  $z$ & H$\delta$ & $\pm$ & O\,{\sc ii} & $\pm$ & type &
Id &  $z$ & H$\delta$ & $\pm$ & O\,{\sc ii} & $\pm$ & type \\
\hline
514 & 0.8398 & $ -2.5 $ & $ 3.0 $ & $ 4.4 $ & $ 3.2 $ & abs  & 1439 & 0.8213 & $ 4.2 $ & $ 2.8 $ & $ 2.3 $ & $ 2.6 $ & E+A? \\ 
564 & 0.8322 & $ 3.2 $ & $ 4.0 $ & $ -39.0 $ & $ 6.8 $ & emi  & 1457 & 0.8420 & $ 2.7 $ & $ 1.2 $ & $ 3.8 $ & $ 1.3 $ & abs  \\ 
573 & 0.8362 & $ 3.7 $ & $ 2.8 $ & $ -1.0 $ & $ 3.1 $ & abs? & 1459 & 0.8461 & $ 5.6 $ & $ 0.9 $ & $ -5.9 $ & $ 1.2 $ & emi? \\ 
661 & 0.8470 & $ 1.0 $ & $ 2.1 $ & $ -5.4 $ & $ 2.2 $ & emi? & 1477 & 0.8274 & $ 1.2 $ & $ 2.6 $ & $ -4.5 $ & $ 3.0 $ & abs? \\ 
675 & 0.8417 & $ 3.2 $ & $ 2.3 $ & $ -1.9 $ & $ 3.1 $ & abs? & 1481 & 0.8258 & $ -0.6 $ & $ 2.0 $ & $ 4.4 $ & $ 2.0 $ & abs  \\ 
696 & 0.8317 & $ 1.3 $ & $ 1.2 $ & $ 1.4 $ & $ 1.5 $ & abs  & 1484 & 0.8318 & $ 1.0 $ & $ 1.0 $ & $ 0.1 $ & $ 1.2 $ & abs  \\ 
696 & 0.8322 & $ 1.4 $ & $ 1.4 $ & $ 0.2 $ & $ 1.5 $ & abs  & 1486 & 0.8332 & $ 0.1 $ & $ 3.4 $ & $ 3.8 $ & $ 4.0 $ & abs  \\ 
703 & 0.8323 & $ -0.5 $ & $ 3.8 $ & $ 1.1 $ & $ 3.7 $ & abs  & 1492 & 0.8280 & $ 1.1 $ & $ 2.4 $ & $ 1.1 $ & $ 2.5 $ & abs  \\ 
710 & 0.8344 & $ 0.4 $ & $ 1.2 $ & $ -1.2 $ & $ 1.5 $ & abs  & 1493 & 0.8334 & $ 2.9 $ & $ 2.9 $ & $ -0.6 $ & $ 3.6 $ & abs? \\ 
742 & 0.8307 & $ 0.3 $ & $ 2.7 $ & $ -2.7 $ & $ 3.0 $ & abs? & 1519 & 0.8301 & $ 1.4 $ & $ 2.1 $ & $ 4.5 $ & $ 2.3 $ & abs  \\ 
777 & 0.8323 & $ 3.7 $ & $ 1.6 $ & $ 2.9 $ & $ 1.9 $ & abs? & 1520 & 0.8310 & $ 0.1 $ & $ 2.7 $ & $ -1.5 $ & $ 2.8 $ & abs  \\ 
793 & 0.8276 & $ 5.3 $ & $ 2.6 $ & $ -2.4 $ & $ 3.1 $ & E+A? & 1567 & 0.8279 & $ 3.4 $ & $ 1.3 $ & $ 2.3 $ & $ 1.4 $ & abs? \\ 
796 & 0.8235 & $ 2.9 $ & $ 1.6 $ & $ 0.8 $ & $ 1.8 $ & abs? & 1583 & 0.8261 & $ 6.9 $ & $ 0.6 $ & $ -0.4 $ & $ 0.7 $ & E+A  \\ 
997 & 0.8389 & $ 0.2 $ & $ 0.7 $ & $ -1.0 $ & $ 0.8 $ & abs  & 1584 & 0.8313 & $ -0.4 $ & $ 1.6 $ & $ -1.1 $ & $ 2.1 $ & abs  \\ 
1035 & 0.8307 & $ 8.1 $ & $ 2.4 $ & $ -4.3 $ & $ 2.9 $ & E+A? & 1585 & 0.8369 & $ 4.4 $ & $ 3.1 $ & $ -3.8 $ & $ 3.3 $ & E+A? \\ 
1088 & 0.8184 & $ 1.0 $ & $ 2.5 $ & $ 3.1 $ & $ 2.1 $ & abs  & 1635 & 0.8363 & $ 1.3 $ & $ 1.7 $ & $ 1.7 $ & $ 2.7 $ & abs  \\ 
1091 & 0.8449 & $ 3.7 $ & $ 1.6 $ & $ -4.6 $ & $ 2.5 $ & abs? & 1639 & 0.8377 & $ 5.4 $ & $ 2.0 $ & $ -8.1 $ & $ 2.3 $ & emi  \\ 
1103 & 0.8222 & $ 5.7 $ & $ 1.9 $ & $ -5.0 $ & $ 1.9 $ & E+A? & 1650 & 0.8297 & $ 8.8 $ & $ 2.1 $ & $ -6.3 $ & $ 2.4 $ & emi? \\ 
1114 & 0.8317 & $ 0.5 $ & $ 2.2 $ & $ 2.1 $ & $ 2.6 $ & abs  & 1650 & 0.8294 & $ 7.1 $ & $ 2.8 $ & $ -3.9 $ & $ 3.1 $ & E+A? \\ 
1119 & 0.8225 & $ -4.2 $ & $ 3.0 $ & $ 4.1 $ & $ 2.3 $ & abs  & 1651 & 0.8241 & $ 0.7 $ & $ 3.5 $ & $ 3.4 $ & $ 3.6 $ & abs? \\ 
1163 & 0.8338 & $ 3.7 $ & $ 1.2 $ & $ -1.1 $ & $ 1.7 $ & abs? & 1656 & 0.8223 & $ 2.0 $ & $ 1.3 $ & $ -0.2 $ & $ 1.5 $ & abs  \\ 
1198 & 0.8313 & $ 1.7 $ & $ 2.4 $ & $ -5.2 $ & $ 2.9 $ & emi? & 1692 & 0.8253 & $ 1.7 $ & $ 2.9 $ & $ -1.6 $ & $ 3.0 $ & abs? \\ 
1209 & 0.8376 & $ 2.5 $ & $ 1.5 $ & $ -0.4 $ & $ 2.0 $ & abs? & 1701 & 0.8317 & $ 1.2 $ & $ 1.7 $ & $ 3.7 $ & $ 2.0 $ & abs  \\ 
1215 & 0.8394 & $ 2.2 $ & $ 1.7 $ & $ -3.4 $ & $ 2.3 $ & abs? & 1733 & 0.8347 & $ 0.5 $ & $ 4.9 $ & $ -23.0 $ & $ 5.6 $ & emi  \\ 
1280 & 0.8372 & $ 0.4 $ & $ 1.3 $ & $ 0.3 $ & $ 1.7 $ & abs  & 1758 & 0.8374 & $ 5.1 $ & $ 2.4 $ & $ 3.0 $ & $ 3.4 $ & E+A? \\ 
1294 & 0.8352 & $ -2.5 $ & $ 1.0 $ & $ 0.3 $ & $ 1.1 $ & abs  & 1760 & 0.8246 & $ 5.6 $ & $ 1.2 $ & $ -1.3 $ & $ 1.4 $ & E+A  \\ 
1298 & 0.8363 & $ 1.5 $ & $ 2.7 $ & $ -5.9 $ & $ 4.3 $ & emi? & 1763 & 0.8390 & $ 3.9 $ & $ 4.1 $ & $ -20.8 $ & $ 4.7 $ & emi  \\ 
1304 & 0.8335 & $ 2.6 $ & $ 1.3 $ & $ 0.4 $ & $ 1.6 $ & abs  & 1801 & 0.8328 & $ 0.6 $ & $ 1.1 $ & $ -51.9 $ & $ 1.6 $ & emi  \\ 
1325 & 0.8311 & $ 2.1 $ & $ 1.0 $ & $ 0.8 $ & $ 1.2 $ & abs  & 1834 & 0.8394 & $ 7.6 $ & $ 1.3 $ & $ -3.1 $ & $ 1.8 $ & E+A  \\ 
1329 & 0.8352 & $ 3.2 $ & $ 2.5 $ & $ -3.5 $ & $ 3.8 $ & abs? & 1855 & 0.8210 & $ 2.7 $ & $ 2.4 $ & $ -17.8 $ & $ 3.8 $ & emi  \\ 
1355 & 0.8355 & $ 1.1 $ & $ 1.3 $ & $ 2.5 $ & $ 1.7 $ & abs  & 1896 & 0.8227 & $ 2.5 $ & $ 3.4 $ & $ -20.6 $ & $ 3.1 $ & emi  \\ 
1359 & 0.8180 & $ 0.9 $ & $ 1.7 $ & $ 1.7 $ & $ 1.6 $ & abs  & 1929 & 0.8400 & $ 2.8 $ & $ 1.5 $ & $ 2.7 $ & $ 1.7 $ & abs? \\ 
1396 & 0.8299 & $ -2.6 $ & $ 2.1 $ & $ 3.5 $ & $ 2.6 $ & abs  & 1942 & 0.8307 & $ 1.8 $ & $ 1.2 $ & $ -2.1 $ & $ 1.5 $ & abs  \\ 
1403 & 0.8132 & $ 2.8 $ & $ 1.9 $ & $ -14.7 $ & $ 2.2 $ & emi  & 1986 & 0.8253 & $ 0.4 $ & $ 1.5 $ & $ -0.1 $ & $ 1.7 $ & abs  \\ 
1405 & 0.8363 & $ 1.2 $ & $ 1.4 $ & $ -0.6 $ & $ 2.0 $ & abs  & 2011 & 0.8413 & $ -4.2 $ & $ 3.2 $ & $ -32.5 $ & $ 3.4 $ & emi  \\ 
1406 & 0.8206 & $ -2.2 $ & $ 1.8 $ & $ -2.8 $ & $ 1.8 $ & abs  & 2130 & 0.8250 & $ 3.1 $ & $ 4.2 $ & $ -10.2 $ & $ 4.5 $ & emi  \\ 
1422 & 0.8330 & $ -2.8 $ & $ 2.9 $ & $ -2.5 $ & $ 2.2 $ & abs  & 2152 & 0.8319 & $ -0.8 $ & $ 1.0 $ & $ -2.8 $ & $ 0.8 $ & abs  \\ 
1430 & 0.8240 & $ 5.4 $ & $ 1.2 $ & $ 0.3 $ & $ 1.3 $ & E+A  & 2174 & 0.8382 & $ -1.6 $ & $ 2.6 $ & $ 2.4 $ & $ 3.3 $ & abs  \\ 
1434 & 0.8248 & $ -1.8 $ & $ 2.3 $ & $ 4.5 $ & $ 2.0 $ & abs  & HES1$^a$ & 0.8286 & $ -11.8 $ & $ 9.2 $ & $ -66.5 $ & $ 14.9 $ & emi  \\ 
1435 & 0.8197 & $ 0.9 $ & $ 2.6 $ & $ 0.8 $ & $ 2.7 $ & abs  & HES2$^a$ & 0.8240 & $ -1.4 $ & $ 5.8 $ & $ -64.0 $ & $ 11.9 $ & emi  \\ 
\hline
\tablecomments{$^a$\,Galaxies unintentially covered by slits pointed at
other objects. These objects have $I \sim 24$ and were not in the photometric
catalog.}
\end{tabular}
\end{center}
\end{table}
\end{small}

\begin{small}
\begin{table}[p]
\begin{center}
{ {\sc TABLE 2} \\
\sc Photometric Data} \\
\vspace{0.1cm}
\begin{tabular}{rcccrrrcccrr}
\hline
\hline
Id & morph & F814W$^T$ & $B_z^{T}$ & $(U-B)_z$ & $\Delta (U-B)_z$ &
Id & morph & F814W$^T$ & $B_z^{T}$ & $(U-B)_z$ & $\Delta (U-B)_z$ \\
\hline
514 &  M/P & 21.83 & 23.00 & $ 0.402 $ & $ -0.022 $ & 1434 &  E/S0 & 21.96 & 23.13 & $ 0.410 $ & $ -0.010 $ \\ 
661 &  Sc & 21.05 & 22.24 & $ -0.064 $ & $ -0.512 $ & 1435 &  E/S0 & 21.87 & 23.04 & $ 0.366 $ & $ -0.057 $ \\ 
675 &  M/P & 21.36 & 22.53 & $ 0.393 $ & $ -0.046 $ & 1439 &  ? & 21.97 & 23.14 & $ 0.461 $ & $ 0.042 $ \\ 
696 &  S0/a & 20.55 & 21.72 & $ 0.467 $ & $ 0.002 $ & 1457 &  M/P & 21.00 & 22.17 & $ 0.448 $ & $ -0.003 $ \\ 
703 &  ? & 21.91 & 23.08 & $ 0.360 $ & $ -0.061 $ & 1459 &  Sc & 20.66 & 21.84 & $ 0.183 $ & $ -0.278 $ \\ 
710 &  E & 20.78 & 21.95 & $ 0.432 $ & $ -0.026 $ & 1477 &  S0/a & 21.57 & 22.74 & $ 0.378 $ & $ -0.055 $ \\ 
777 &  S0 & 20.99 & 22.16 & $ 0.431 $ & $ -0.020 $ & 1478 &  Sa & 21.37 & 22.54 & $ 0.496 $ & $ 0.057 $ \\ 
793 &  M/P & 21.14 & 22.31 & $ 0.365 $ & $ -0.082 $ & 1481 &  S0 & 21.95 & 23.12 & $ 0.373 $ & $ -0.047 $ \\ 
796 &  S0 & 21.80 & 22.97 & $ 0.413 $ & $ -0.012 $ & 1484 &  E & 19.53 & 20.70 & $ 0.496 $ & $ -0.002 $ \\ 
997 &  M/P & 20.29 & 21.47 & $ 0.343 $ & $ -0.130 $ & 1486 &  Sa & 22.00 & 23.17 & $ 0.534 $ & $ 0.116 $ \\ 
1024 &  ? & 21.75 & 22.95 & $ -0.194 $ & $ -0.619 $ & 1492 &  ? & 21.95 & 23.12 & $ 0.440 $ & $ 0.020 $ \\ 
1035 &  M/P & 22.34 & 23.53 & $ -0.122 $ & $ -0.528 $ & 1493 &  Sb & 22.30 & 23.47 & $ 0.454 $ & $ 0.045 $ \\ 
1039 &  Sa & 21.31 & 22.49 & $ 0.333 $ & $ -0.108 $ & 1519 &  S0/a & 21.21 & 22.38 & $ 0.454 $ & $ 0.010 $ \\ 
1043 &  E & 20.87 & 22.04 & $ 0.461 $ & $ 0.006 $ & 1520 &  E/S0 & 21.57 & 22.74 & $ 0.442 $ & $ 0.009 $ \\ 
1086 &  S0 & 21.65 & 22.82 & $ 0.424 $ & $ -0.006 $ & 1567 &  E & 21.04 & 22.21 & $ 0.494 $ & $ 0.044 $ \\ 
1088 &  E & 21.92 & 23.09 & $ 0.423 $ & $ 0.002 $ & 1583 &  M/P & 20.50 & 21.68 & $ 0.318 $ & $ -0.149 $ \\ 
1091 &  S0 & 22.05 & 23.23 & $ 0.215 $ & $ -0.201 $ & 1584 &  E & 20.76 & 21.93 & $ 0.483 $ & $ 0.024 $ \\ 
1103 &  E/S0 & 21.95 & 23.14 & $ 0.059 $ & $ -0.361 $ & 1585 &  Sc & 22.09 & 23.28 & $ 0.004 $ & $ -0.411 $ \\ 
1114 &  S0/a & 21.44 & 22.61 & $ 0.410 $ & $ -0.027 $ & 1635 &  S0/a & 21.77 & 22.94 & $ 0.467 $ & $ 0.041 $ \\ 
1119 &  S0 & 21.85 & 23.02 & $ 0.442 $ & $ 0.019 $ & 1639 &  Sb & 21.59 & 22.77 & $ 0.200 $ & $ -0.232 $ \\ 
1163 &  M/P & 20.49 & 21.66 & $ 0.438 $ & $ -0.029 $ & 1650 &  Sd & 21.59 & 22.78 & $ 0.056 $ & $ -0.375 $ \\ 
1198 &  Sa & 22.17 & 23.36 & $ 0.031 $ & $ -0.381 $ & 1651 &  E/S0 & 22.17 & 23.34 & $ 0.463 $ & $ 0.050 $ \\ 
1209 &  S0 & 21.41 & 22.58 & $ 0.504 $ & $ 0.066 $ & 1655 &  E & 20.94 & 22.12 & $ 0.252 $ & $ -0.200 $ \\ 
1215 &  Sa & 21.09 & 22.26 & $ 0.421 $ & $ -0.027 $ & 1656 &  E/S0 & 21.06 & 22.23 & $ 0.448 $ & $ -0.001 $ \\ 
1280 &  E & 20.74 & 21.91 & $ 0.482 $ & $ 0.022 $ & 1692 &  M/P & 21.68 & 22.86 & $ 0.313 $ & $ -0.115 $ \\ 
1294 &  E/S0 & 20.60 & 21.77 & $ 0.472 $ & $ 0.008 $ & 1701 &  S0 & 21.10 & 22.27 & $ 0.447 $ & $ -0.001 $ \\ 
1298 &  Sa & 22.11 & 23.28 & $ 0.469 $ & $ 0.054 $ & 1733 &  Sc & 22.07 & 23.26 & $ -0.036 $ & $ -0.452 $ \\ 
1304 &  Sb & 21.28 & 22.45 & $ 0.396 $ & $ -0.046 $ & 1758 &  S0/a & 21.93 & 23.11 & $ 0.354 $ & $ -0.066 $ \\ 
1305 &  Sb & 22.04 & 23.22 & $ 0.335 $ & $ -0.082 $ & 1760 &  M/P & 20.29 & 21.46 & $ 0.369 $ & $ -0.105 $ \\ 
1325 &  E & 19.88 & 21.05 & $ 0.478 $ & $ -0.009 $ & 1763 &  Sb & 22.71 & 23.90 & $ -0.086 $ & $ -0.481 $ \\ 
1329 &  Sb & 20.87 & 22.04 & $ 0.393 $ & $ -0.062 $ & 1801 &  M/P & 20.39 & 21.60 & $ -0.536 $ & $ -1.005 $ \\ 
1340 &  M/P & 21.06 & 22.23 & $ 0.472 $ & $ 0.023 $ & 1834 &  Sb & 20.98 & 22.16 & $ 0.326 $ & $ -0.125 $ \\ 
1354 &  Sb & 21.51 & 22.68 & $ 0.424 $ & $ -0.010 $ & 1896 &  Sc & 21.12 & 22.32 & $ -0.210 $ & $ -0.656 $ \\ 
1355 &  E/S0 & 21.38 & 22.55 & $ 0.455 $ & $ 0.017 $ & 1929 &  E & 21.18 & 22.35 & $ 0.410 $ & $ -0.035 $ \\ 
1359 &  E/S0 & 21.09 & 22.26 & $ 0.457 $ & $ 0.010 $ & 1942 &  E/S0 & 20.98 & 22.15 & $ 0.457 $ & $ 0.006 $ \\ 
1396 &  S0/a & 21.60 & 22.77 & $ 0.486 $ & $ 0.055 $ & 1986 &  Sb & 21.30 & 22.47 & $ 0.524 $ & $ 0.083 $ \\ 
1403 &  Sc & 20.14 & 21.33 & $ -0.037 $ & $ -0.515 $ & 2011 &  M/P & 22.13 & 23.33 & $ -0.271 $ & $ -0.685 $ \\ 
1405 &  E & 20.03 & 21.20 & $ 0.478 $ & $ -0.004 $ & 2130 &  Sd & 21.84 & 23.03 & $ -0.001 $ & $ -0.424 $ \\ 
1406 &  E & 21.42 & 22.59 & $ 0.398 $ & $ -0.040 $ & 2174 &  S0/a & 21.82 & 22.99 & $ 0.372 $ & $ -0.052 $ \\ 
1422 &  Sc & 21.03 & 22.23 & $ -0.230 $ & $ -0.679 $ & HES1 &  Sd & 23.21 & 24.23 & $ -0.194 $ & $ -0.752 $ \\ 
1430 &  Sb & 20.36 & 21.54 & $ 0.348 $ & $ -0.123 $ & & & & & & \\ 
\hline
\end{tabular}
\end{center}
\end{table}
\end{small}

\end{document}